\documentclass[usenatbib,iop,numberedappendix]{aeb_emulateapj_2010}
\usepackage{amsmath}
\usepackage{amssymb}

\usepackage{natbib}

\usepackage[usenames,dvips]{color}

\def\s{{\rm s}} %...........seconds
\def\ns{{\rm n}\s} %........nanoseconds
\def\Ms{{\rm M}\s} %........megaseconds
\def\yr{{\rm yr}} %.........years
\def\Myr{{\rm M}\yr} %......megayears
\def\Gyr{{\rm G}\yr} %......gigayears
\def\hr{{\rm hr}}

\def\MHz{{\rm MHz}} %.......Megahertz
\def\GHz{{\rm GHz}} %.......Gigahertz
\def\m{{\rm m}} %...........meters
\def\mm{{\rm m}\m} %........millimeters
\def\cm{{\rm c}\m} %........centimeters
\def\km{{\rm k}\m} %........kilometers
\def\pc{{\rm pc}} %.........parsecs
\def\kpc{{\rm k}\pc} %......kiloparsecs
\def\Mpc{{\rm M}\pc} %......megaparsecs
\def\Ms{M_\odot} %...........solar masses

\def\Jy{{\rm Jy}} %.........Janksy (1e-26 J/sHzm^2 = 1e-23 erg/sHzcm^2)
\def\mJy{{\rm m}\Jy} %......milliJansky

\def\AU{{\rm AU}} %.........astronomical unit

\def\mas{{\rm mas}} %.......milli-arcseconds
\def\muas{\mu{\rm as}} %....micro-arcseconds
\def\rad{{\rm rad}} %.......radians

\renewcommand{\d}{{d}}
\newcommand{\e}{{e}}
\def\Ms{{M_\odot}}

\newcommand\bmath[1] {\mbox{\boldmath$\rm #1$}}

\def\bx{\bmath{x}}
\def\bu{\bmath{u}}
\def\bhu{\bmath{\hat{u}}}
\def\bd{\bmath{d}}
\def\balpha{\bmath{\alpha}}
\def\bdelta{\bmath{\delta}}

\def\HtO{${\rm H}_2{\rm O}$}

\begin{document}

\title{
Localizing Sagittarius A* and M87 on Microarcsecond Scales with Millimeter VLBI
}

\author{
Avery E.~Broderick\altaffilmark{1},~
Abraham Loeb\altaffilmark{2} and
Mark J.~Reid\altaffilmark{3}
}
\altaffiltext{1}{Canadian Institute for Theoretical Astrophysics, 60 St.~George St., Toronto, ON M5S 3H8, Canada; aeb@cita.utoronto.ca}
\altaffiltext{2}{Institute for Theory and Computation, Harvard University, Center for Astrophysics, 60 Garden St., Cambridge, MA 02138.}
\altaffiltext{3}{Harvard-Smithsonian Center for Astrophysics, 60 Garden St., Cambridge, MA 02138.}

\shorttitle{Localizing Sgr A* and M87 on Microarcsecond Scales}
\shortauthors{Broderick, Loeb \& Reid}

\begin{abstract}
With the advent of the {\em Event Horizon Telescope} (EHT), a
millimeter/sub-millimeter very-long baseline interferometer (VLBI), it
has become possible to image a handful of black holes with sub-horizon
resolutions.  However, these images do not translate into
microarcsecond absolute positions due to the lack of absolute phase
information when an external phase reference is not used.
Due to the short atmospheric coherence time at these wavelengths,
nodding between the source and phase reference is impractical.
However, here we suggest an alternative scheme which makes use of the
fact that many of the VLBI stations within the EHT are arrays in their
own right.
With this we show that it should be possible to absolutely position
the supermassive black holes at the centers of the Milky Way (Sgr A*)
and M87 relative to nearby objects with precisions of roughly
$1\,\muas$.
This is sufficient to detect the perturbations to Sgr A*'s position
resulting from interactions with the stars and stellar-mass black
holes in the Galactic cusp on year timescales, and severely constrain
the astrophysically relevant parameter space for an orbiting
intermediate mass black hole, implicated in some mechanisms for
producing the young massive stars in the Galactic center.
For M87, it allows the registering of millimeter images, in which the
black hole may be identified by its silhouette against
nearby emission, and existing larger scale radio images, eliminating
present ambiguities in the nature of the radio core and inclination,
opening angle, and source of the radio jet.
\end{abstract}


\maketitle

\section{Introduction} \label{I}

The {\em Event Horizon Telescope} (EHT), a proposed Earth-sized array of
existing millimeter and sub-millimeter observatories, promises to
provide microarcsecond imaging resolutions via very-long baseline
interferometry (VLBI) \citep{Doel_etal:2010}.  This is sufficient to
resolve the horizons of a handful of supermassive black holes,
including Sagittarius A* (Sgr A*) and M87.  Already, a mm-VLBI array
consisting of three stations has probed sub-horizon scale structure
in Sgr A* at $1.3\,\mm$ wavelengths
\citep[][]{Doel_etal:08,Fish_etal:10}. These have
been used to test fundamental physics, providing 
direct evidence for the existence of an event horizon surrounding the
supermassive black hole at the center of the Milky Way
\citep{Brod-Loeb-Nara:09}, as well as study the accretion flow
surround Sgr A*, giving an estimate for the black hole spin
\citep{Brod_etal:09,Huan-Taka-Shen:09,Mosc_etal:09,Dext-Agol-Frag-McKi:10,Brod_etal:10}.

The existing observations have made use of the
{James Clerk Maxwell Telescope} (JCMT) and the
{Sub-Millimeter Array} (SMA), located atop Mauna Kea in Hawaii,
the {Combined Array for Research in Millimeter-wave Astronomy} (CARMA)
in Cedar Flat, California, and the {Arizona Radio Observatory Sub-Millimeter Telescope}
(SMT) on Mount Graham, Arizona, giving a maximum baseline of
$4500\,\km$, oriented predominantly east-west.  Potential future
stations include sites in Chile (Atacama Pathfinder EXperiment,
Atacama Submillimeter Telescope and Atacama Large Millimeter Array;
APEX, ASTE, and ALMA, respectively), Mexico (Large Millimeter
Telescope; LMT), the South Pole (South Pole Telescope; SPT),
and the IRAM telescopes in Spain (Pico Veleta; PV) and France (Plateau
de Bure; PdB).  Among these the longest baselines are
$\simeq1.2\times10^4\,\km$, corresponding to a synthesized beam of
$30\,\muas$ at $230\,\GHz$ ($1.3\,\mm$) and $20\,\muas$ at $345\,\GHz$
($0.87\,\mm$).

In addition to including new baselines, improvements in sensitivity are
actively being pursued, both by pushing towards larger bandwidths and
collecting areas.  Using the JCMT, SMT and a single $10\,\m$ antenna
at CARMA, \citet{Fish_etal:10} was able to measure $0.1\,\Jy$
visibilities on long baselines.  Efforts already underway to increase
the bandwidth to $4\,\GHz$ will improve that by a factor of two.
Phasing together the multiple telescopes on Mauna Kea \citep[SMA, JCMT \&
CSO;][]{Wein:08} and many antenna at CARMA will improve this further,
increasing sensitivity by a factor of 3 at some sites, making it
possible to detect sources considerably dimmer than Sgr A* and M87.
This will make it possible to contemplate doing 
astrometry on microarcsecond scales using the EHT.

Since the advent of self-calibration, phased-reference VLBI
observations have only been necessary in imaging observations for
low-brightness sources.  Nevertheless, because the use of closure phases
discards an over-all phase calibration, phase-referenced observations
provide the unique capability to do relative astrometry between
distinct sources, e.g., positioning a target relative to distant
background quasars.  In the case of the Galactic Center, $7\,\mm$-VLBI
observations using the Very-Long Baseline Array have already paid
substantial scientific dividends: placing tight constraints upon the
motion of Sgr A*
\citep{Reid-Read-Verm-Treu:99,Reid-Brun:04,Reid_etal:08}, providing 
independent parallax measurements of our distance from the Galactic
center \citep{Reid_etal:09}, and registering
the radio and infrared reference frames of the central $0.1\,\pc$
\citep{Ment_etal:97,Reid_etal:03,Reid_etal:07}.  All of these make
specific use of background quasars to locate sources within the
Galactic center to sub-milliarcsecond precisions.

Moving towards shorter wavelengths is challenging for a number of
reasons, including the typically rapidly declining source flux
densities and shortened atmospheric coherence time ($10\,\s$) and
angle (isoplanatic angle, $\sim3^\circ$, see Appendix \ref{app:MACL}).
Nevertheless,
there are strong scientific
motivations for attempting to do so.  Chief among these is that at
mm-wavelengths the detection of the black hole silhouette
unambiguously identifies and locates the supermassive black holes in
Sgr A* and M87.  Registering the mm and cm-VLBI reference frames then
provides a means to determine where the black hole is relative to
larger morphological features.  Microarcsecond astrometry also allows
the direct detection of the jitter induced by the interaction of the
supermassive black holes with the retinue of compact objects in their
vicinity, providing a probe of the dark component of the stellar cusp
surrounding the central object.  Finally, absolute astrometry makes it
possible to study short-timescale morphological changes in Sgr A* and
M87.

Here we describe some of the the scientific motivations for
phase-referenced mm-VLBI in more detail, giving estimates of
characteristic time scales and precisions needed to detect the
relevant phenomena.  Since the particular questions that are
accessible are different in Sgr A* and M87, we will address each
separately.  Section \ref{sec:PRMV} describes a possible scheme by
which mm-VLBI may be accomplished, including some of the critical
limitations that the short wavelength imposes.  Sections
\ref{sec:SGRA} and \ref{sec:M87} provide the motivations as well as
candidate phase references that meet the requirements set out in
Section \ref{sec:PRMV}.  Finally, concluding remarks may be found in
Section \ref{sec:C}.

\section{Phase Referenced Millimeter-VLBI} \label{sec:PRMV}

With the advent of self-calibration, interferometric radio observations
became capable of reconstructing the vast majority of complex
visibility phase information in a given observation, correcting for
the nearly all of the station-dependent errors due to propagation
through the atmosphere.  These methods are, however, insensitive to an
overall phase shift of the complex visibilities, including to a
translational uncertainty in the image.  This precludes VLBI
observations from measuring the absolute position of radio features
at the incredible precision that it can image them.

This situation may be ameliorated with the use of a phase reference,
which provides a fixed point against which to measure the relative
location of the source.  For this purpose, background quasars
are frequently used, due to their small intrinsic angular size,
high brightness and exceedingly small transverse velocities.  In
practice, the small angular size and fixed location of the phase
reference provides an easily modeled system with which to reconstruct
with high accuracy the phase errors at each VLBI station, fulfilling an
analogous role as guide stars in optical and infrared adaptive optics
systems\footnote{In this context, when we refer to ``phase-referenced
  mm-VLBI'' we are not describing a scheme for performing mm-VLBI
  observations on dim sources.  Rather, we are discussing a means to
  assess the relative positions of sources that are sufficiently
  bright to be detected independently.}.

Any potential phase reference must be sufficiently close to the target
that it suffers strongly correlated phase distortions, i.e., it must
lie within the isoplanatic patch.  At mm-wavelengths this
depends upon the particular site, typically requiring
separations $\lesssim2^\circ$--$3^\circ$, though we will discuss this in 
more detail below (see also Appendix \ref{app:MACL}).  Similarly, the
phase reference must be detectable
on timescales comparable to the atmospheric coherence time, roughly
$10\,\s$ at mm-wavelengths.  In practice, these impose 
severe limitations on phase-referenced mm-VLBI.  While a handful of
potential phase references may exist within $2^\circ$ of the target
source, none lie within the beam widths of individual EHT stations,
which range from $6''$ (LMT) to $54''$ (SMA).  As a consequence,
observations of the target source and phase reference with a single
dish requires repointing the telescopes.  However, this must be 
accomplished well within the atmospheric coherence time, leaving
sufficient time to detect both the source and the phase reference,
something that is presently not possible.  

However, the EHT is unique in that many of the VLBI stations are
interferometers in their own right, i.e., contain many independent
antennas.  As a consequence, at these stations it is possible to
observe both the target source and phase reference simultaneously
using a subset of the available antennas for each
\citep[the {\em Paired Antennas} approach described in Chapter 13.2 of
][]{Thom-Mora-Swen:01}.  The remaining instrumental components of the
complex gains change slowly in comparison to the atmospheric
propagation effects, and thus may be measured by periodically
switching antennas from one subset to the other.  This is identical to
the approach used by \citet{Coun_etal:74} to detect the gravitational
deflection of quasars via VLBI, and similar to the principles
underlying VERA \citep{Kawa-Sasa-Mana:00,Honm-Kawa-Sasa:00}.

\begin{deluxetable}{llcccc}\tabletypesize{\small}
\tablecaption{EHT Baseline Sensitivities for Multi-Antenna Stations \label{tab:fl}}
\tablehead{
\colhead{Realization} &
\colhead{Baseline} &
\colhead{$F_{230} (\mJy)$} &
\colhead{$F_{345} (\mJy)$} &
\colhead{$F_{230,0}^M (\Jy)$} &
\colhead{$F_{345,0}^M (\Jy)$}
}
\startdata
EHT-I & ${\rm H^J}$--${\rm C^1}$ & $41$  & $63$  & $1.3$   & $1.7$\\
-- & ${\rm H^J}$--${\rm A^1}$    & $19$  & $34$  & $0.61$  & $0.91$\\
-- & ${\rm C^1}$--${\rm A^1}$    & $27$  & $45$  & $0.88$  & $1.2$\\
-- & ${\rm P^1}$--${\rm A^1}$    & $17$  & $36$  & $0.56$  & $0.94$\\
EHT-C & ${\rm H^a}$--${\rm C^8}$ & $10$  & $16$  & $0.33$  & $0.42$\\
-- & ${\rm H^a}$--${\rm A^{10}}$ & $3.9$ & $7.1$ & $0.13$  & $0.19$\\
-- & ${\rm C^7}$--${\rm A^{10}}$ & $3.3$ & $5.5$ & $0.11$  & $0.15$\\
-- & ${\rm P^6}$--${\rm A^{10}}$ & $2.2$ & $4.6$ & $0.072$ & $0.12$
\enddata
\end{deluxetable}

Stations consisting of arrays are some of the most sensitive in the EHT,
providing the most permissive flux limits upon potential phase
references.  Potential stations for this purpose include Hawaii,
CARMA, ALMA and Plateau de Bure.  The limiting sensitivities at
$230\,\GHz$ and $345\,\GHz$ on baselines that include these
stations at both ends, assuming a bandwidth of $4\,\GHz$ and
$10\,\s$ integration times, are listed in Table \ref{tab:fl}.  We
include sensitivity estimates for two possible realization:
\begin{description}
\item[EHT-I] Using only the JCMT in Hawaii (${\rm H^J}$), a single $10.4\,\m$
  antenna at CARMA (${\rm C^1}$), $12\,\m$ antenna at ALMA
  (${\rm A^1}$), and a single $15\,\m$ antenna at Plateau de Bure 
  (${\rm P^1}$).
\item[EHT-C] Using the phased Hawaii station (including the SMA, JCMT and
  CSO, ${\rm H^a}$), 8 CARMA antennas (6 $10.4\,\m$ and 2 $6.1\,\m$
  antennas, ${\rm C^8}$), 10 $12\,\m$ ALMA antennas (${\rm A^{10}}$)
  and all 6 $15\,\m$ Plateau de Bure antennas (${\rm P^6}$).
\end{description}
The first of these indicate what is possible with limited number of
antenna, similar to what is possible for initial version of the EHT,
available in the near future.  The second corresponds to the maximum
achievable sensitivities using the complete EHT though employing only
existing facilities.  To estimate these flux limits we used the 
system-equivalent flux densities reported in \citet{Doel_etal:09} 

In Appendix \ref{app:MACL} we estimate the atmospheric limits upon the
accuracy with which angular separation may be measured.  In principle,
this is also limited by source structure and variability, though
imaging with the EHT should ameliorate these (indeed, as described in
Section \ref{sec:SgrA-IDAEF} the characterization of
source variability is one of the motivations for high-precision
astrometry), and instrumental phase noise.  Atmospheric limitations
come in two forms: that due to the small-scale, rapidly varying
structures that limit the coherence timescale, and that due to
large-scale, slowly evolving features that induce anomalous phase
delays resulting from the slighly different zenith angle, $z$, of the target
and reference sources.  As a result, the mitigation strategies for
these differ substantially.  Generally, we find that sub-beam
precision is almost certainly achievable, with $\muas$ accuracy likely
possible.  Note, however, that this accuracy is not always required.
For example, the precision with which it is possible to register
mm-VLBI and existing cm-VLBI images is necessarily limited by the
resolution of the latter, which is roughly $0.1\,\mas$.

As described in Appendix \ref{app:MACL}, the short timescale of the
rapid component implies that hundreds of realizations of the
atmospheric turbulence are encountered in a typical observing night
(roughly $2\,\hr$ of on-source integration time).
For phase references within
$1.4^\circ$ the atmospheric conditions will permit beam-scale
($30\,\muas$) astrometry on $10\,\s$ intervals at $230\,\GHz$.  Over
$10\,\min$, the integration times used in \citet{Doel_etal:09} and
\citet{Fish_etal:10}, this can be reduced to $4\,\muas$ scales.  In a
$2\,\hr$ period this may be reduced to $1\,\muas$.  Thus, in practice,
this is unlikely to be the dominant contribution to the astrometric
uncertainty.

At $7\,\mm$, the slowly varying, large-scale component dominates the
astrometric uncertanties \citep{Reid-Read-Verm-Treu:99},
producing typical unmodeled time delays of $0.1\,\ns$ for sources
separated by $1^\circ$ in azimuth.  This corresponds to an uncertainty
on the order of $0.1\,\mas$, i.e., three beam widths at $1.3\,\mm$.
However, there are a variety of schemes that may be employed to reduce
this significantly.  All of these seek to do so by exploiting the
large spatial and temporal scales of the atmospheric fluctuations
responsible for this component of the astrometric uncertainty.

The first is to implement better atmospheric models.  The atmospheric
model of \citet{Niel:96} already accounts for the seasonal variations,
generating daily estimates for the zenith phase delay, and thus does
model the very-long timescale fluctuations.  Improvements
over this model come in a variety of forms, including the structure
of the atmosphere assumed \citep{Boeh-Neil-Treg-Schu:06} and the
inclusion of real-time weather modeling to produce accurate maps on
$6\,\hr$ timescales \citep{Boeh-Werl-Schu:06}.  These promise to
reduce the size of the unmodeled phase perturbations by factors of a
few.

The second is to periodically measure the group delays directly.  This
can be done using sets of known bright quasars, which are assumed not
to move, so-called geodetic blocks, and has already been done in
astrometric studies of the Galactic Center
\citep{Reid-Brun:04,Reid_etal:09}.  Alternatively, this can be done
using Global Positioning Systems (GPS) and Global Navigation Satellite
Systems (GNSS) \citep[see, e.g.,][]{Byun-BarS:09}.  Already, GPS/GNSS
is used to produce estimates for the zenith path delay with a
time resolution of $5\,\min$ and precision of roughly $3\,\mm$ at sites
comparable to those that will be employed by the EHT.  This alone
would be sufficient to improve the astrometric uncertainty due to the
slow component by an order of magnitude, giving an accuracy of roughly
$10\,\muas$.

If three or more viable references are located in the vicinity of the
target\footnote{Note that because the atmospheric fluctuations that generate the
  slowly-varying phase perturbations are necessarily large, additional
  reference sources are not subject to the same constraints that limit the
  reference--target separation.  Since the distortion in the
  separation grows linearly with zenith separation, larger separations
  between reference sources result in larger relative phase
  perturbations due the large-scale fluctuations.
  However, the uncertainty associated with the rapid variations also
  grow nearly linearly, and thus the fractional precision with which the
  slowly-varying phase perturbations can be measured grows only slowly.
  Nevertheless, the number of bright references grows as separation
  squared.  The separations are fundamentally limited by the region over
  which the large-scale atmospheric fluctuations can be accurately
  modeled.}, a third strategy is to calibrate the unknown large-scale
phase delays directly.  This can be done by asserting that the
true positions of the references are fixed, and then using a procedure
similar to that described in Appendix \ref{app:MRAC} to estimate the
site specific phase delays by comparing the observed and expected
reference separations \citep[see also][]{Foma:05}.  This
assumes that the references are not 
variable themselves, though even at $\muas$ resolutions distant
quasars are unlikely to exhibit structural variability on hour
timescales.  In principle, as shown in the appendix, the
accuracy with which this can be done is comparable to the precision
limit associated with the short-timescale variability, roughly
$1\,\muas$.

Finally, the slow variations can be averaged down by combining
observations over many days.  The upper limit upon the time over which
the observations can be combined is dictated by the time-scale of
variations that are of interest.  Thus, e.g., fluctuations in the
position of Sgr A* on year-long timescales allows averaging over
$\sim3$ months, corresponding to $\sim10^2$ independent realizations of
the large-scale fluctuations, and thus a reduction in the uncertainty
due to the slow-component of an additional order of magnitude.

In practice a combination of these mitigation strategies are likely to
be required.  Nevertheless, there is clearly a good reason to believe
that $\muas$ accuracies are achievable over long times via improved
atmospheric modeling, direct phase delay measurement and time
averaging.  Furthermore, it may be possible to reach these accuracies over
much shorter times using multiple phase references, though this is
more speculative.

In the following sections we discuss some of the science made possible
by phase-referenced mm-VLBI observations, and suggest possible phase
references to be used for these purposes.  Since many of the
scientific motivations are different for Sgr A* and M87, the
observing strategies also differ.  Therefore, we address these sources
separately.

\section{Sagittarius A*} \label{sec:SGRA}

Associated with the supermassive black hole located at the center of
the Milky Way, Sgr A* is the primary target for the EHT.  Presently,
the best estimates of the mass ($M$) and distance ($D$) of Sgr A* come
from the observations of orbiting stars (the so-called ``S-stars'').
These have yielded $M=4.3\pm0.5\times10^6\,\Ms$ and $D=8.3\pm0.4\,\kpc$,
respectively, where both include the systematic uncertainties
\citep{Ghez_etal:08,Gill_etal:09a,Gill_etal:09b}.  This results in an
angular size for the apparent horizon of
$53\pm2\,\muas$\footnote{Because the current mass and distance
  measurements are strongly correlated, with mass scaling roughly as
  $M\propto D^{1.8}$ \citep{Ghez_etal:08}, the angular size of Sgr A*
  is constrianed much more strongly than $M$ or $D$ alone.},
the largest for any known black hole.  The mass of Sgr A* is
necessarily confined to within the periapse of nearby stars, giving a
maximum radius of roughly $10^2\,\AU\simeq3\times10^3GM/c^2$.

\subsection{Scientific Objectives}
A variety of astrometric studies of Sgr A* have already been proposed
and carried out.  As a consequence of the large extinction towards the
Galactic center, these necessarily are confined to the radio and
near-infrared (NIR).  Nevertheless, a wide variety of objectives have
been and are being pursued, including placing independent constraints
upon the mass of the central supermassive black hole
\citep{Reid_etal:99,Reid-Brun:04,Reid-Brun:05} and detecting
structural variations during radio and NIR flares
\citep{Reid_etal:08,Bart_etal:09}.  A generic feature of these is the
need for high astrometric precision.  Here we describe how these
efforts may be complemented or improved upon using the $\muas$
precision afforded by mm-VLBI, as well as suggest additional
scientific motivations for developing a mm-VLBI astrometry capability.

\subsubsection{Constraints upon the Proper Motion of Sgr A*}
The current hierarchical paradigm for black hole formation and active
galactic nuclei (AGN) fueling is fundamentally dynamic, imparting large
velocities to the central supermassive black holes.  However, on
timescales short in comparison to the time between galactic mergers
this motion is dissipated by collective interactions between the
supermassive black hole and the stars in the stellar bulge, i.e.,
dynamical friction \citep{Chat-Hern-Loeb:02}.  Thus we expect the
black hole at the center of the Milky Way to be nearly at rest,
located at the bottom of the Galactic potential well.  However, because of
stochastic interactions with nearby stars, it will not be completely
stationary and, if the surrounding stellar core is well described by a
Plummer sphere, will execute random motions (perpendicular to the Galactic
plane, say) with variance
\begin{equation}
\langle v_\perp^2 \rangle = \frac{2}{9} \frac{G M_{\rm tot}}{a} \frac{m_\star}{M}
\end{equation}
where $M_{\rm tot}$ is the total mass of the nuclear stellar cluster,
$m_\star$ is the characteristic mass of a star and $a$ is the core
radius of the Plummer sphere \citep{Chat-Hern-Loeb:02,Reid-Brun:04}.
Thus, given an estimate of the stellar content within some radius,
$R$, it is possible to constrain the the mass of Sgr A* given an upper
limit upon its motion:
\begin{equation}
M > \frac{M_R}{1 + (9/2)\left\{ \langle v_\perp^2 \rangle a R^3\big/
\left[G\left(R^2+a^2\right)^{3/2} m_\star\right]\right\}}\,,
\end{equation}
\citep{Chat-Hern-Loeb:02,Reid-Brun:04}.  Note that this constraint is
independent of those obtained from measurements of the orbits of stars
immediately surrounding Sgr A*.

High-precision astrometric studies of Sgr A* have already been
performed at 7mm using the Very Long Baseline Array (VLBA),
constraining $\langle v_\perp^2 \rangle < 2\,\km \s^{-1}$,
corresponding to $M>2\times10^6\,\Ms$.  They have also measured the
apparent velocity parallel to the Galactic plane, giving
$v_\parallel\simeq250\pm16\,\km\,\s^{-1}$ (assuming a distance of
$8.3\,\kpc$ to Sgr A*), which is presumably dominated by the
contribution from the Sun's orbit in the Galaxy.  The VLBA
observations have a nominal astrometric 
precision of $10^2\,\muas$, and extend over a period of more than a
decade.  During this time they see no evidence for fluctuations,
implying that either the motion is secular or has a period
significantly exceeding $\sim20\,\yr$.  Note that the natural period
for oscillations about the center of the average collective Galactic
potential is dependent upon the radius ($R_{\rm inf}$) of the region
of influence, and roughly
$7.4\times10^3\left(R_{\rm inf}/0.3\,\pc\right)^{-3/2}\,\yr$, much
larger than any observational period of interest.

Empirical estimates for the magnitude of the secular motion of Sgr A*
are presently limited by the accuracy of individual radio observations.
In principle, the uncertainty in the perpendicular component of the
peculiar motion of the Sun provides a fundamental limit upon efforts
to measure the {\em secular} motion of Sgr A*.  This is presently
estimated to be $7.25\pm0.37\,\km\,\s^{-1}$ relative to the local
standard of rest \citep{Scho-Binn-Dehn:10}.  An additional uncertainty
may arise from Galactic precession, which can be as high as
$0.6\,\km\,\s^{-1}$ in the extreme case in which the Milky Way
``flips'' in a Hubble time. These corresponds to an angular deviation
of $9.4\,\muas\,\yr^{-1}$ and $15\,\muas\,\yr^{-1}$, respectively,
both much smaller than the limits imposed by the accuracy of existing
radio observations.  Therefore, considerable improvement in the
astrometric accuracy is possible in principle.

\subsubsection{Detection of the Evolved Component of the Cusp} \label{sec:cusp}
In addition to a secular component due the Sun's orbit and a potential
displacement from the center of the Galactic potential,
the velocity of the central supermassive black hole has a fluctuating
component due to random forces applied by fluctuations in the
surrounding retinue of stars and their compact remnants.  In contrast
to the secular motion, detecting a fluctuating component is made much
simpler due to the variations in the velocity, which clearly
differentiates it from contributions due to the solar motion.  With
$\muas$ astrometric precision, detecting this fluctuating component on
scales of a few years should be possible.

It is from this fluctuating component that the mass estimates
described in the preceding section is ultimately derived.  However, it
is also possible to probe the spatial and mass distribution of the
stars and their remnants in the immediate vicinity of Sgr A*.
Understanding the properties of these objects is of intrinsic
interest; how stars form and evolve in the Galactic center is
presently poorly understood, characterized by the so-called ``paradox
of youth'' \citep{Ghez_etal:03}.  However, it is also important as an exemplar.
Detecting the stellar component would provide a direct means with
which to estimate the rates of stellar disruptions and the subsequent
contribution to the growth of AGN.

Furthermore, orbiting populations of stellar-mass black holes are
expected to play a pivotal role in present and future gravitational
wave searches.  Stellar-mass black hole binary coalescence rates for
the Laser Interferometer Gravitational Wave Observatory (LIGO) are
strongly affected by the concentration of the stellar-mass black holes
in galactic nuclei \citep{OLea-Kocs-Loeb:09}.
Similarly, the event rate of extreme mass-ratio inspirals (EMRIs), 
one of the primary science drivers for the Laser Interferometer Space
Antenna (LISA) \citep[see, e.g.][]{Mill_etal:09,Shut_etal:09}, is
strongly dependent upon the density and mass function of stellar-mass
black holes near the supermassive black hole \citep{Sigu-Rees:97,Gair_etal:04,Hopm-Alex:05,Hopm-Alex:06,Gair-Tang-Volo:10,Amar-Pret:10}. 
We note that the environment of Sgr A* is particularly relevant to
LISA, since it is comparable in mass to the systems that will dominate
the LISA frequency band.  In both cases, empirical estimates of the
black hole distribution in Sgr A* can substantially reduce the
uncertainties in the inferred merger rates.

Here we estimate the precision necessary to directly observe the
motion induced upon Sgr A* by the nuclear stellar and black hole
populations.  To do this we assume a fiducial set of distributions for
these objects, and consider both optimistic and pessimistic deviations
from these.  In all cases the natural frequency of oscillation about
the minimum of the Galactic potential and the damping timescale due to
dynamical friction with the stellar bulge ($\omega_0$ and $\gamma$ in
Appendix \ref{app:PJPS}) are relevant only at much larger timescales
than the observationally motivated subset we consider (i.e.,
sub-century).

The stellar mass density surrounding Sgr A* has been inferred from
infrared studies, finding that the total stellar mass density is well
approximated by a broken power law:
\begin{equation}
\rho_\star(r) \simeq 1.7\pm0.8\times10^6\left(\frac{r}{0.22\,\pc}\right)^{-\gamma}
\,\Ms\,\pc^{-3}
\,,
\label{eq:rhoS}
\end{equation}
where $\gamma=1.2$ for $r<0.22\,\pc$ and $\gamma=1.75$ for $r>0.22\,\pc$
\citep{Scho_etal:07}.  This is well constrained observationally for
$r>10^{-2}\,\pc$, though there is considerable uncertainty regarding
the structure of the core on smaller scales \citep{Merr:10}.  In
addition there is evidence for a preponderance of young
stars in the vicinity of the supermassive black hole
\citep{Buch_etal:09,Do_etal:09,Do_etal:10}.  This biases the stellar mass
function in the Galactic center towards higher masses.  Nevertheless,
here we make the conservative choice of adopting a stellar mass
function indicative of the Galactic disk.  Specifically, we employ the
log-normal mass function for low-mass stars determined by
\citet{Cove_etal:08}, smoothly joining a Salpeter high-mass tail,
i.e., $\phi(m,r)\propto m^{-2.3}$ above $m\simeq0.6\,\Ms$, and
truncated at $10^2\,\Ms$.  For our purposes here, it is enough to note
that the average stellar mass and it's square are $m_\star=0.75\,\Ms$ and
$\mu_\star^2\simeq 13.7\,\Ms^2$, respectively.  Note that the Salpeter
mass function immediately implies that $\mu_\star^2(r)$ is determined
by the maximum mass and is generally much larger than $m_\star^2$.

Less well constrained is the contribution from stellar remnants.  Large
numbers of stellar-mass black holes are expected to collect via
dynamical friction in the vicinity of the central supermassive black
hole \citep{OLea-Kocs-Loeb:09,Alex-Hopm:09}.  A number of authors
have estimated the number of stellar-mass black holes in the Galactic
center, finding that roughly $2\times10^4$ are expected to exist
within the inner $\pc$
\citep{Morr:93,Mira-Goul:00,Hopm-Alex:06,Frei-Amar-Kalo:06,Alex-Hopm:09}.
The distribution of these remnants is less well understood;
predictions ranging from flat cores to power-law cusps with indexes as
extreme as $-2.75$ can be found in the literature
\citep[see, e.g.,][]{Alex-Hopm:09,Merr:10}, though cusps 
with an indexes near $-2$ are most common.  Thus we adopt a fiducial
model for the black hole component given by
\begin{equation}
\rho_\bullet = \frac{m_\bullet N_{\pc}}{4\pi} \left(\frac{r}{\pc}\right)^{-2} \,\Ms\,\pc^{-3}\,,
\label{eq:rhoBH}
\end{equation}
with $m_\bullet=10 \Ms$ and $N_{\pc}$ is the number of black holes
in the inner $\pc$.

In Appendix \ref{app:PJPS} we derive an approximate power
spectrum for the displacement of the central
supermassive black hole, $P_{x,\omega}$, for an arbitrary number
density of objects, $n(r)$, and a potentially radially dependent mass
function, $\phi(m,r)$:
\begin{equation} 
P_{x,\omega}
\simeq
\omega^{-4}
\int_0^\infty \d r \, 
\frac{2(2\pi)^{3/2}}{3} \frac{G^2\mu^2}{r^2} \frac{n}{\Omega_k}
\e^{-\omega^2/2\Omega_k^2}\,,
\end{equation}
with $\Omega_k$ given by Equation (\ref{eq:Ok}).  In this $\mu$ is
the RMS mass at a given radius (thus weighted towards the more massive
objects).  At this point, there are some general features of note.
First, since $P_{x,\omega}$ depends upon the object mass function
through $\mu^2 n \sim \mu^2 \rho/m$ alone, when the same mass
density is present, perhaps unsurprisingly, more massive
stars/remnants result in proportionally larger displacements.  Second,
if $\mu^2$ is fixed, $n(r)\propto r^{-\nu}$ with $\nu>1/2$, and the
motions of the stars/remnants is dominated by the central supermassive
black hole, we have
\begin{equation}
P_{x,\omega} \simeq \frac{2 (2\pi)^{3/2}}{2\nu-1}
\frac{G^2\mu^2}{r_{\rm min}}  \frac{n_{\rm max}}{\Omega_{k,\rm max}}
 \omega^{-4} 
\quad{\rm for}\quad
\omega < \Omega_{k,\rm max}\,,
\end{equation}
i.e, $P_{x,\omega}$ is dominated by contributions from those objects
nearest the central black hole and has a characteristic frequency
dependence independent of the stellar distribution power law index.
Thus, it is possible to make robust predictions for the scale of the
jitter of Sgr A*, dependent primarily upon the maximum density and
typical masses of stellar-mass objects in Sgr A*'s immediate vicinity.

\begin{figure}
\begin{center}
\includegraphics[width=\columnwidth]{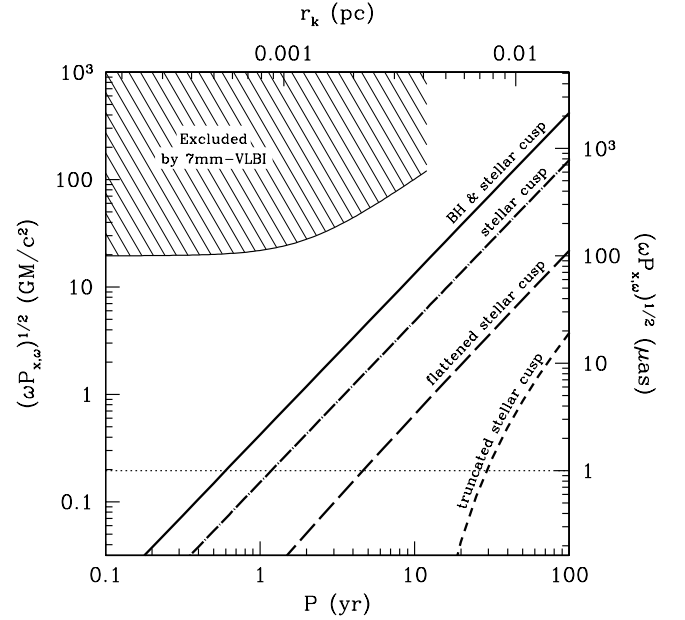}
\end{center}
\caption{Characteristic displacements as a function of time-scale for
  when stellar mass black holes and stars are considered (solid), only
  the stellar component is considered (dot-dashed), when the stellar
  component is assumed to form a core of size $0.01\,\pc$
  (long-dashed), and when all components are truncated inside of
  $0.01\,\pc$ (short-dashed).  The region excluded by $7\,\mm$ observations
  with the VLBA is shown by the hatched area \citep{Reid-Brun:05}.
  Along the left and right axes the displacements are shown in units
  of black hole mass and angle.  Along the top axis the radius with
  Keplerian period equal to the time period is listed, providing some
  sense for the location of the objects that are dominating the
  fluctuations.}\label{fig:disp}
\end{figure}

Figure \ref{fig:disp} shows the characteristic displacements, given by
$\sqrt{\omega P_{x,\omega}}$, for a handful of possible
stellar/remnant core models.  In all but one the power spectrum
follows it's asymptotic $\propto \omega^{-4}$ form (corresponding to
$\omega^{-3/2}$ in the figure), despite the fact that $n(r)$ is not a
single power law and $\mu^2(r)$ ranges from $13.7\,\Ms^2$ at large
radii to $10^2\,\Ms^2$ at small radii.  What varies amongst the power
spectra is primarily the normalization.

The solid line shows our most
optimistic case, with the stellar and black hole components given by
Equations (\ref{eq:rhoS}) and (\ref{eq:rhoBH}), respectively,
truncated at the innermost stable circular orbit (ISCO) of the central
black hole, located at roughly $1.2\times10^{-6}\,\pc$.  In this case
the displacement spectrum is dominated by the black hole component,
and produces fluctuations in excess of a $\muas$ on $0.5\,\yr$
timescales.  If, however, we ignore the remnant component altogether,
the stellar component given by Equation (\ref{eq:rhoS}) produces the
dot-dashed line, implying that observable displacements will occur on
$1\,\yr$ timescales.  In both cases the typical displacements on
year-long timescales are comparable to those estimated in
\citet{Reid-Brun:04} (which find typical velocities of
$0.3\,\km\,\s^{-1}$ and $0.07\,\km\,\s^{-1}$, producing angular
displacements of $8\,\muas$ and $2\,\muas$ due to the remnant and
stellar components, respectively\footnote{In \citet{Reid-Brun:04} the
  displacements due to remnants was twice as large as that reported
  here due to their assumption of twice as much mass in the black hole
  cusp.}).
This presumes that we are justified in
extrapolating the observed stellar density all the way down to the
ISCO, over four orders of magnitude.  If instead we force the stellar
density to saturate at $0.01\,\pc$, the smallest scale for which the
stellar distribution has been measured, we obtain the long-dashed
line.  Even in this case observable motions are present on sub-decade
timescales.  Finally, if we make the extreme pessimistic choice of
truncating the stellar and black hole distributions at $0.01\,\pc$ we
obtain the short-dashed spectrum, the curve of which is indicative of
the lack of objects with sufficiently short periods to generate the
short-timescale fluctuations.

\subsubsection{Searching for a Massive Binary Companion}

The hierarchical paradigm of galaxy formation also implies the
presence of supermassive black hole binaries at the centers of some
galaxies.  Initially the supermassive black holes sink toward the center
of the combined galaxy as a result of dynamical friction due to
stars in the reassembled bulge and drag imparted by gas funneled into
the galactic center by the merger.  However, this process becomes
inefficient when the binary separation shrinks to a few $\pc$, at
which point the stellar encounters that carry away the orbital angular
momentum become rare.  This is before gravitation radiation is able to
drive the system to merger within a Hubble time, resulting in the
so-called ``final parsec'' problem \citep{Bege-Blan-Rees:80,Roos:81}.

A number of solutions to the final parsec problem have been suggested,
including triaxial bulges \citep{Merr-Poon:04}, massive perturbers
\citep{Pere-Alex:08}, and interactions with the merger enhanced gas
density \citep{Esca_etal:04,Esca_etal:05}.  Nevertheless, in some
fraction of systems long-lived supermassive black hole binaries are
expected
\citep{Bege-Blan-Rees:80,Yu-Trem:03,Milo-Merr:03,Gual-Merr:09}. For
this reason searches for supermassive binary companions have been
performed in a variety of systems, primarily using spectroscopic
methods
\citep{Bogd-Erac-Sigu:08,Boro-Laue:09,Deca_etal:10,Liu-Shen-Stra-Gree:10}.

In the specific context of Sgr A*, massive companions and perturbers
have been implicated for a very different reason: as a solution to the
paradox of youth, i.e., the presence of the massive, and therefore
young, stars orbiting Sgr A* \citep{Ghez_etal:03}.  Formation of the
``S-stars'' (typically B-type stars) in situ is
thought to be precluded by the strong tidal forces from Sgr A*.  This
has lead to the development of models in which the stars are formed
further out, where the restrictions are less severe, and then
transported into Sgr A*'s vicinity \citep{Gerh:01}.  However, it is only
possible to reach the small radii observed if the S-stars were members
of a very tightly bound cluster, requiring a heretofore undetected
massive component, the prime candidate therefore being an intermediate
mass black hole
\citep{McMi-Port:03,Hans-Milo:03,Kim-Fige-Morr:04,Levi-Wu-Thom:05,Fuji-Iwas-Funa-Maki:09,Merr-Gual-Mikk:09,Fuji-Iwas-Funa-Maki:10}.
That is, the transport models imply the presence of a massive object
located at a distance comparable to that of the S-star orbits, roughly
$10^{-2}\,\pc$, from Sgr A*.

Thus, in addition to the stochastic buffeting from stellar-mass black
holes and stars, Sgr A* may also move as a consequence of a massive
binary companion.  Whether or not this is observable astrometrically
depends upon the orbital parameters, mass of the perturber and
timescale of observations.  Some set of these is already ruled out by
the lack of an oscillatory signal in the $7\,\mm$ proper motion
studies \citep{Reid_etal:99,Reid-Brun:04,Reid-Brun:05}.  If we assume
circular orbits with periods short in comparison to the observation
time\footnote{For decade-long observations, this is true for
  $a\lesssim0.01\,\pc$.  Above $0.01\,\pc$ the orbital periods exceed
  $30\,\yr$, and thus only a small fraction of the orbit would have
  been observed, decreasing the ability to probe these cases.
  Furthermore, at these distances the velocity is nearly constant, and
  thus degenerate with that due to the solar motion.}, 
these rule out companion masses
\begin{equation}
m \gtrsim 8\times10^4 \left(\frac{a}{10^2\,\AU}\right)^{-1}\,\Ms\,,
\label{eq:CMc}
\end{equation}
\citep[see also][]{Hans-Milo:03}.  In the presence of eccentricity
this expression is not a hard limit, since fortuitous orientations can
result in small projected displacements.

\begin{figure}
\begin{center}
\includegraphics[width=\columnwidth]{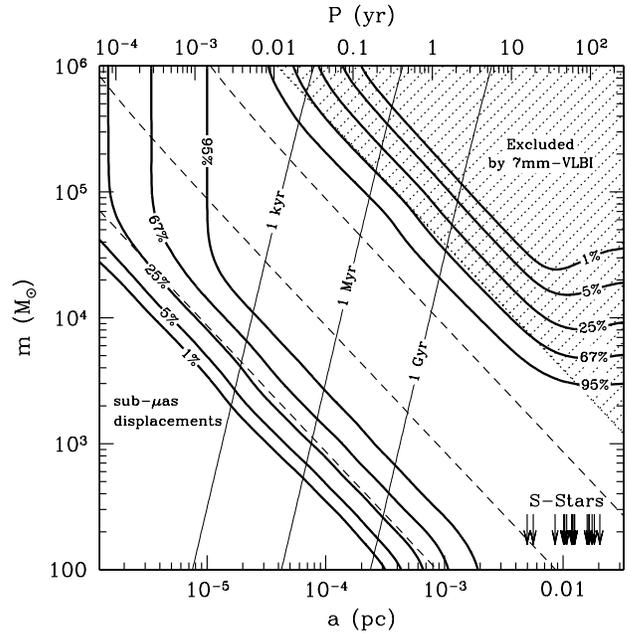}
\end{center}
\caption{Non-excluded region of the companion mass--orbital
  separation parameter space that can be probed by astrometry with
  $\muas$ precisions \citep[cf.][]{Hans-Milo:03}.  Thick solid lines
  show (from outside in) 
  locations in the parameter space with 1\%, 5\%, 25\%, 67\%, and 95\%
  probabilities of not having been previously detected by previous
  astrometric studies at $7\,\mm$ with the VLBA (violated in the upper
  right) while exhibiting projected position shifts larger than a
  $\muas$ at some point over the next decade (violated in the lower
  left).  The dotted hatched area shows the approximate region
  described in the text that is ruled out by $7\,\mm$ VLBA studies.
  For reference, lines of constant maximum angular deviation are shown
  for circular orbits by the dashed lines, corresponding to
  $1\,\muas$ (leftmost), $10\,\muas$, and $100\,\muas$ (rightmost).
  Gravitational merger timescales for circular orbits are shown by the
  thin solid lines, ranging from $10^{-3}\,\yr$ (leftmost) to
  $1\,\Gyr$ (rightmost).  The period of the orbit (in the
  test particle limit) is shown along the top axis.  Finally, the
  semi-major axes of the S-stars are indicated by the arrows.}\label{fig:mc}
\end{figure}

Adopting a simple two-body model (i.e., ignoring the many-body effects
of the stellar/remnant cusp) we estimate the likelihood of a given
set of orbital parameters both satisfying the existing astrometric
limits and executing deviations larger than $1\,\muas$ over the next
decade. This provides some idea of the relevant parameter space that
$\muas$-astrometry can exclude in the near future.  Figure
\ref{fig:mc} shows this likelihood, marginalized over orbital
orientation and eccentricity (assuming a flat prior) as a function of
orbital separation and companion mass.  As expected, large masses and
orbital separations are already ruled out by \citet{Reid-Brun:05}
(specifically, by requiring that the displacements be less than that
observed perpendicular to the Galactic plane).  While including
eccentric orbits shift the location of this constraint, Equation
(\ref{eq:CMc}) follows the contours of constant likelihood, falling
nearly upon the 67\% confidence level.  At larger separations, the
orbital period substantially exceeds the timescale over which the
$7\,\mm$ VLBA observations of Sgr A* have been performed, and their
diagnostic ability decreases.  At small masses and orbital separations
the companion induces only sub-$\muas$ shifts in the position of Sgr
A*, and is thus not likely to be detectable by the EHT.  Even when the
companion mass is sufficiently large, highly eccentric orbits pass
within the ISCO, restricting the allowed parameter space further.
Additionally, massive companions on small orbits merge rapidly, though
an intermediate mass black hole associated with the S-stars would
necessarily have entered the region in the past $\Myr$.  At
large separations the fluctuations due to stellar-mass black holes
described in Section \ref{sec:cusp} provide a source of additional
noise (not shown in Figure \ref{fig:mc}).  However, this is confined
to the extreme lower right of Figure \ref{fig:mc} even in the most
optimistic cusp models.  Regardless, for reference, the dashed lines
in Figure \ref{fig:mc} show the lower limits achievable with
$1\,\muas$, $10\,\muas$, and $100\,\muas$ astrometric precision,
assuming circular orbits.  Thus, high-precision astrometry with the
EHT should prove particularly effective for detecting massive
companions to Sgr A*.

\subsubsection{Imaging Dynamical Accretion/Ejection Features} \label{sec:SgrA-IDAEF}

The astrometric observations we have discussed thus far have focused upon
detection of changes in the physical location of the supermassive black
hole.  However, astrometry is also sensitive to changes in the
structure of the source.
The observation of radio, NIR and X-ray flares in
Sgr A*, with characteristic timescales ranging from $\sim20\,\min$ to
hours, is strong evidence for substantial variability near the black
hole horizon itself
\citep{Genz_etal:03,Ecka_etal:06,Meye_etal:06,Porq_etal:08,Marr_etal:08}.
Highly variable, turbulent, magnetized 
plasmas flows are a generic component of both accretion and jet models
for Sgr A*'s emission, and thus it is not surprising that transient
features arise near the black hole with dynamical times comparable to
the orbital timescale at the ISCO.  Sgr A*'s flares are therefore
naturally explained within the context of transient structural changes
to the emitting region, e.g., orbiting hot spots generated by magnetic
reconnection events or shocks and confined within a disk, expanding
blobs of energetic material buoyantly ejected from a disk, or hot
spots launched outwards in a jet.  If this is indeed the case, flares
should produce observable temporary fluctuations in Sgr A*'s location.

As new bright features emerge or move within the source
the centroid of the image shifts, following the bright structures.  The
motion of the image centroid associated with orbiting hot spots, in
the context of Sgr A*'s NIR flares, has been discussed at length
\citep{Brod-Loeb:05,Brod-Loeb:06b}, and is the primary motivation
for the GRAVITY instrument being developed for the Very Large
Telescope Interferometer \citep[see, e.g.][]{Bart_etal:09}.  At
$7\,\mm$ the opacity of the accretion flow precludes the observation
of spots orbiting near the ISCO.  Nevertheless, it is possible to
detect the motion associated with bright spots orbiting more than
$30GM/c^2$ from the black hole using the VLBA \citep{Reid_etal:08}.
The EHT bridges the two regimes, and should access the innermost
portions of the accretion flow with sufficient resolution to resolve
any transient spots.

Detecting structural changes does not necessarily require absolute
astrometry.  If the image is not dominated by the dynamical
feature, approaches based upon closure quantities and/or the evolution
of the source polarization are capable of detecting changes in the
shape and orientation of the emission region
\citep{Doel_etal:09,Fish_etal:09}.  Alternatively, if the image
uniformly brightens or expands, closure techniques can characterize
the evolution of the emitting region.  However, when a single
transient feature dominates the emission, as has been observed to be
the case for bright flares in the NIR, and this feature evolves
structurally, absolute astrometry provides the only method to
unambiguously detect the spatial variations.

More importantly, since the dynamical timescale in Sgr A* is
$\sim10\,\min$, the assumptions underlying Earth-rotation aperture synthesis
are explicitly violated during flares.  Specifically, we are no longer
justified in assuming that the underlying source structure remains
fixed throughout the night.  For the EHT this is especially limiting;
even when all existing and planned mm and sub-mm facilities are phased
together, like the VLBA, the EHT will only be able to provide sparse
instantaneous coverage of the $u$--$v$ plane.  With a sufficiently
well specified model for the variability in hand, it is possible to
analyze the visibilities directly, eliminating the need to produce
images altogether.  However, in the absence of such a model phase
referencing provides the only means by which to relate visibilities at
different times, producing approximate images of the time-averaged
intensity.

Furthermore, in the presence of a single dominant spot, phase
referencing may be required to remove the fundamental spot-disk
degeneracy.  That is, in the absence of an identifying feature that
locates the black hole (e.g., the silhouette of the horizon, which may
not be visible in the instantaneous images), fitting the
self-calibrated visibilities alone will not be sufficient to
distinguish a moving spot (either orbiting, infalling or outflowing)
about a fixed quiescent disk and an artificially fixed spot with an
artificially moving quiescent disk.  Of course, absolute astrometry
naturally resolves this ambiguity.

To address these issues, it is necessary to achieve sub-beam
($\lesssim 30\,\muas$) astrometric precision on the dynamical
timescale of Sgr A* ($\sim20\,\min$).  As long as a sufficiently
strong phase reference is within $\sim1^\circ$, this should be possible
with the EHT.

\subsection{Possible Phase References}

Phase reference candidates must satisfy a number of constraints.
Chief among these are brightness and compactness; the reference must
be detectable in mm-VLBI observations.  Most of what we describe in
this section are efforts to characterize these properties.  However,
in addition the structure of the phase reference and Sgr A* must be
stationary over the timescale of the experiment.  In the case of the
latter this is assisted by direct imaging, which can unambiguously
identify the location of the black hole via it's silhouette against
the nearby emitting material.  For the phase references the case for
stability is less clear, and ultimately, this will have to be
empirically established.  Here we will assume that this can be done.
For quasars this is well justified, being located at extraordinary
distances.  For masers this may not be, limiting their utility to
short-term (days or weeks) uses only.  Using multiple phase references
provides a means to test this assumption and potentially mitigate
violations.

%\newpage
\subsubsection{Quasars}
\begin{figure*}
\begin{center}
\includegraphics[width=\textwidth]{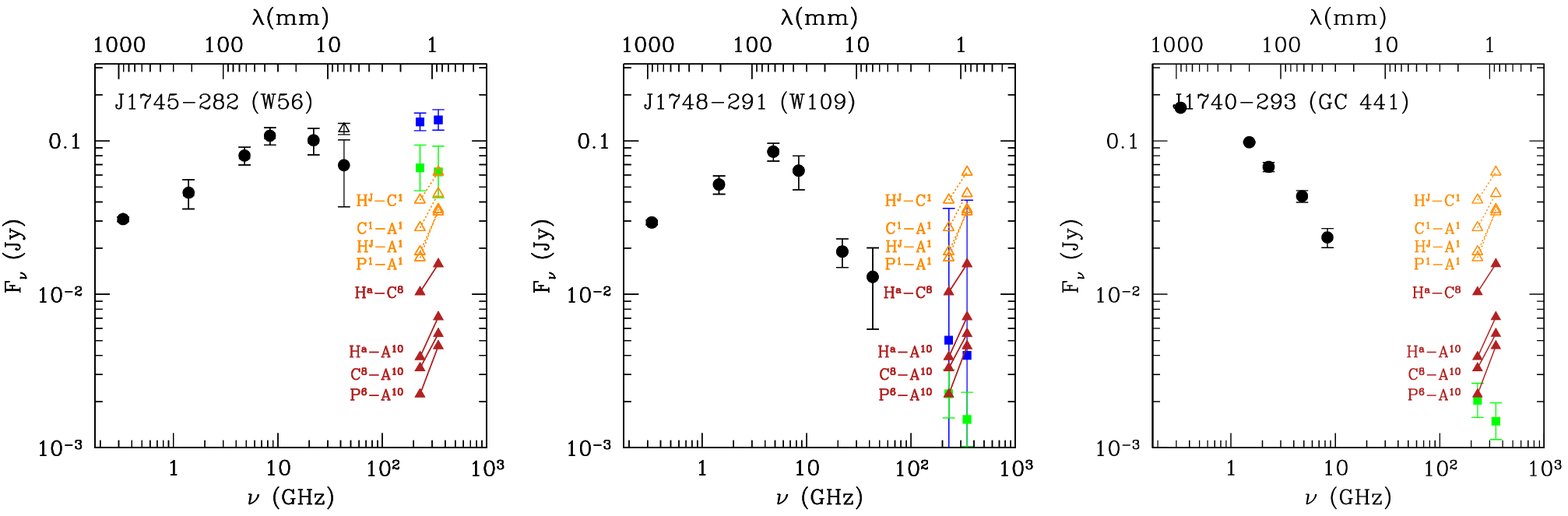}
\end{center}
\caption{SEDs of radio-bright quasars within $1^\circ$ of Sgr A*
  considered by \citet{Bowe-Back-Sram:01}.  The flux densities shown
  by the black circles were taken from the literature and are not
  contemporaneous \citep{Isaa-Wout-Habi:80,Zoon_etal:90,Bowe_etal:99,Bowe-Back-Sram:01,Bowe-Falc-Saul-Back:02,Nord_etal:04,Roy-Rao-Subr:05}.  Thus the error bars indicate the larger of the
  intrinsic measurement uncertainty or the variability amongst
  multiple data points at a given frequency.  Green squares
  show the inferred flux densities at $230\,\GHz$ and $345\,\GHz$.
  For J1745-282 the $43\,\GHz$ flux measurement is highly variable,
  due presumably to observing conditions, and thus we also show the
  inferred high-frequency flux densities associated with the brightest
  $43\,\GHz$ measurement by the blue squares.  For J1748-291 the blue
  squares show the mm flux densities inferred from VLBA observations
  alone.  For reference the sensitivities of baselines consisting of
  stations with multiple telescopes at both ends are shown at
  $230\,\GHz$ and $345\,\GHz$, by the open-orange and filled-red
  triangles (see the text for details).} \label{fig:SgrASED}
\end{figure*}

The quasars J1745-282 (W56), J1748-291 (W109), and J1740-293 (GC 441)
are located within $1^\circ$ of Sgr A*, and have been
used previously
as phase references for proper motion studies of Sgr A* with the VLA
\citep{Back-Sram:99}.
The first two have been successfully used in VLBI observations at
wavelengths of $3.5\,\cm$ and above
\citep{Reid_etal:99,Bowe-Back-Sram:01,Reid-Brun:04}.  We show the
spectral energy densities (SEDs) of these sources are in Figure
\ref{fig:SgrASED}.  These are constructed from the literature and are
therefore not contemporaneous, varying over nearly 30 years
\citep{Isaa-Wout-Habi:80,Zoon_etal:90,Bowe_etal:99,Bowe-Back-Sram:01,Bowe-Falc-Saul-Back:02,Nord_etal:04,Roy-Rao-Subr:05}.
As a consequence, the uncertainties reflect the larger of either that
associated with the intrinsic measurement or that due to the
variability implied by the multi-epoch observations. 

All of these are optically thin above $43\,\GHz$.  At frequencies
above this transition we assume the SED is characterized by a
power-law, which we extrapolate to obtain single-dish flux densities
at $230\,\GHz$ and $345\,\GHz$.  Inferring the magnitudes of
visibilities on long baselines then requires some assumption about the
source structure; here we make the extreme simplifying assumption
that the candidate references are point sources.  Therefore, we
compare the extrapolated flux densities directly to those presented in
Table \ref{tab:fl}.  However, these may be considerable overestimates
if substantial structure is not present on scales smaller than
$\sim10^2\,\muas$ in these objects.  While the point source
approximation may not be justified
(e.g., the flat high-frequency spectrum of J1745-282 suggests the
presence of a compact optically thick component, and thus extended
structure), the observed sizes of the candidate references at longer
wavelengths are consistent with those of point sources, scatter
broadened by the intervening interstellar electrons
\footnote{The interstellar electron scattering that broadens images of
  the Galactic Center is subdominant at $\mm$ wavelengths.  Since the
  size of the scattering kernel, well described by an asymmetric
  gaussian, scales as $\lambda^2$, this is not true at longer
  wavelengths, where scattering necessarily induces correlated flux
  losses.}.

The steep spectrum of J1740-293, shown in the bottom panel of Figure
\ref{fig:SgrASED}, predicts fluxes well below the limits 
along any baseline at both $230\,\GHz$ and $345\,\GHz$.  This is
unsurprising since J1740-293 is already too dim at $43\,\GHz$ to be
employed as a phase reference \citep{Reid-Read-Verm-Treu:99}.  Thus,
while we include it here for completeness, it is almost certainly
excluded as a potential phase references at shorter wavelengths as
well.  By comparison, both J1748-291 and J1745-282 are used as phase
references for proper motion studies of Sgr A*
\citep{Reid-Read-Verm-Treu:99,Reid-Brun:04,Reid_etal:08}.

J1748-291 transitions into the optically thin regime between
$1.4\,\GHz$ and $8.4\,\GHz$, above which the SED appears well
characterized by a power-law.  As seen in the middle panel of Figure
\ref{fig:SgrASED}, there is some uncertainty in the mm flux densities,
deriving from the considerable uncertainty in the $22\,\GHz$ and
$43\,\GHz$ fluxes.  This arises in part due to the lower resolution of
the VLA observations at $8.4\,\GHz$ in comparison to the VLBA
measurements at higher frequencies.  While these agree within their
respective uncertainties, including only VLBA data points produces a
flatter spectrum on average, and thus higher mm flux densities.
Nevertheless, it is clear that J1748-291 is unlikely to be
sufficiently bright to be used as a phase reference by the EHT-I.  For
the EHT-C, it is still too dim to
be detected on the Hawaii--CARMA baseline, though may achieve
marginal detections on baselines including more than 10 ALMA antenna.
For this reason we consider J1748-291 to be a marginal case.

In contrast, J1745-282 is almost certainly bright enough to be used as
a phase reference at $230\,\GHz$ for both the EHT-I and EHT-C, and
$345\,\GHz$ in the case of the latter.  The spectral 
index of J1745-282 is poorly constrained at present since the SED
transitions to the optically thin regime near or above $8.4\GHz$.
More importantly, the $43\,\GHz$ flux estimates are limited by
atmospheric transmission effects, which bias the estimates towards
lower values.  As a result, the highest observed flux
values may be better estimates of the high-frequency luminosity of
J1745-282 than the average value.  Therefore, in the interest of
completeness we show both, and their associated estimates, in the left
panel of Figure \ref{fig:SgrASED}.  In the former case the spectrum
remains inverted at high frequencies, implying flux densities of
$\sim0.1\,\Jy$ at mm wavelengths, well in excess of the various
detection limits.  In the latter case, the flux densities are more
modest, $\sim30\,\mJy$, and comparable to the limits for the EHT-I.
Nevertheless, both estimates imply that for the EHT-C J1745-282 is a
strong mm-VLBI phase reference candidate.

\subsubsection{Masers}
Masers provide an attractive alternative due to their small intrinsic
size and large brightness temperature.  Many molecular transitions
exist that can produce masers at high frequencies, and those at mm and
sub-mm wavelengths have only begun to be investigated \citep{Hump:07}.
Known sites of maser emission in the Galactic center include AGB
stars \citep[e.g., IRS 7, ][]{Ment_etal:97,Reid_etal:03,Reid_etal:07}
and nearby star formation regions \citep[e.g., Sgr B2, ][]{Qin_etal:08,Reid_etal:09,Casw-Bree-Elli:10}.

The SiO masers associated with AGB stars are believed to
originate within 4--5 stellar radii ($\sim8\,\AU$) of the star itself.
However, typical AGB stellar radii are $\sim300R_\odot$, implying an
angular scale for the masing region of roughly $1\,\mas$
\citep{Reid-Ment:07}.  As a consequences, many of these sources will
almost certainly be resolved out by mm-VLBI observations, eliminating
them as candidate phase references.

In contrast, \HtO~masers have been observed to have typical extents
comparable to an AU within star forming regions, corresponding to an
angular scale $\sim10^2\,\muas$ \citep{Reid-Mora:81}, and \HtO~maser
spots an order of magnitude smaller have been observed surrounding
late-type stars \citep{Imai_etal:97}.  At $22\,\GHz$, \HtO~masers have
already been used in VLBI astrometry observations within the Galactic
center.  \citet{Reid_etal:09} report the detection of a number of
strong maser sources in the Sgr B2 complex with typical sizes of
$\sim0.3\,\mas$.  This is much smaller than the $\sim2\,\mas$
scatter-broadened size of a point source co-located with Sgr A* at
$22\,\GHz$, and thus it is unclear how much of the observed typical
maser spot size arises due to interstellar electron scattering.  As a
consequence the Sgr B2 maser spot observations may be consistent with
much smaller intrinsic sizes.  High-frequency \HtO~maser transitions
at $325\,\GHz$ and $380\,\GHz$ have been observed with fluxes similar
to that of associated $22\,\GHz$ \HtO~masers surrounding late-type
stars.

As with quasars, the need to detect the masing region within the
atmospheric coherence time produces a flux limit, below which the
maser is not a viable phase reference candidate.  Unlike continuum
sources, however, the line-like nature of the maser prevents the EHT
from leveraging large bandwidths to increase sensitivity.  Rather, the
signal-to-noise ratio is maximized when the observing bandwidth matches
the maser line width, which for typical velocity widths of $\Delta
V\sim 5\,\km\,\s^{-1}$ are roughly $4\,\MHz$ at $233\,\GHz$.  The
corresponding detection threshold is then dependent upon the velocity
width, with
\begin{equation}
F_\nu^M = F_{\nu,0}^M \left(\frac{\Delta V}{5\,\km\,\s^{-1}}\right)^{-1/2}
\end{equation}
for which the coefficients $F_{\nu,0}^M$ are given in Table
\ref{tab:fl}.  These are typically on the order of $1\,\Jy$ for the
EHT-I and $0.3\,\Jy$ for the EHT-C.  By comparison, these are much
smaller than the typical flux of $40\,\Jy$ reported for the Sgr B2M
spot observed at $22\,\GHz$ by \citet{Reid_etal:09}.

\section{M87} \label{sec:M87}

M87 is well known as a prototypical AGN, exhibiting a powerful
relativistic jet.  Recent analyses of its surrounding bulge have
produced mass and distance estimates for the central supermassive
black hole of $6.6\pm0.4\times10^9\,\Ms$ and $17.9\,\Mpc$,
respectively \citep{Gebh-Thom:09,Gebh_etal:11}.  As a consequence, the angular size
of M87's horizon is roughly $70\%$ that of Sgr A*'s, and thus M87 is
an excellent second target for the EHT.

\subsection{Scientific Objectives} \label{sec:M87:SO}

The scientific objectives that can be addressed with short-wavelength
astrometric observations in M87 are qualitatively different from those
described for Sgr A* for two reasons.
First, M87's much larger mass results in
correspondingly substantially larger dynamical scales.  Unlike Sgr A*,
M87 evolves slowly, with the period at the ISCO of
$\gtrsim 5\,{\rm days}$, and thus using Earth-rotation aperture synthesis is
well justified even for epochs of large variability.  Similarly the time
and mass scales for massive companions are increased by
$\sim1.5\times10^3$.  Thus, decade long observations can only search
for objects within $\sim1.4\times10^2 GM/c^2$, and more massive than
$\sim10^7\,\Ms$.  However, such a supermassive black hole binary will
merge in less than $5\,\Myr$ due to gravitational radiation alone, and
is thus exceedingly unlikely.  Finally, fluctuations in M87's position
driven by stellar-mass black holes are essentially undetectable as a
result of both, the increased timescale and the much larger mass of
the central object.

Second, in contrast to Sgr A*, M87 powers an ultrarelativistic jet.
For this reason M87 has already been subject of a number of
high-resolution imaging efforts
\citep{Juno-Bire-Livi:99,Ly-Walk-Wrob:04,Kric_etal:06,Kova-List-Homa-Kell:07,Ly-Walk-Juno:07,Walk-Ly-Juno-Hard:08}.
Nevertheless, the relationship between the observed jet emission and
the black hole remains unclear \citep[see, e.g.][]{Mars_etal:08}.  By
resolving the horizon, the EHT will unambiguously identify the central
black hole.  Registering the mm/sub-mm and radio VLBI images will
therefore allow the conclusive determination of the location of the
black hole relative to the larger-scale radio features.  Note that to
achieve this, $\muas$ precision is not required; the ability to
position the radio VLBI and mm/sub-mm images is fundamentally limited
by the astrometric precision at longer wavelengths, roughly
$30\,\muas$.  Positioning the black hole in this manner provides
immediate insight into a variety of aspects of the relationship
between the central supermassive black hole and the surrounding
larger-scale emission, with associated implications for jet formation. 
In the remainder of this section we expand upon what can be learned by
relating the mm and radio images.

\subsubsection{Locating the Counter Jet with Implications for Inclination and Acceleration}
\citet{Ly-Walk-Juno:07} and \citet{Kova-List-Homa-Kell:07} report tantalizing evidence for the presence
of a counter jet in M87.  This arises from stacking multiple
self-calibrated images, though it is probably not a calibration
artifact.  However, the interpretation of this feature as a counter
jet hinges upon the identification of the core as the black hole
location.  In the absence of phase referenced VLBA observations of
M87 showing symmetric outflowing motions, this remains an assumption.

The presence of a counter jet places severe constraints upon the
inclination angle, $\Theta$, and acceleration of M87's jet.  Making
the simplifying assumptions that the jet and counter jet are
symmetric, with constant spectral indexes
($S_\nu\propto\nu^{-\alpha}$), and only moving radially with Lorentz
factor $\Gamma$, the observed brightness ratio is approximately
\begin{equation}
R_B
\equiv
\frac{I_J}{I_{CJ}}
\simeq
\left(\frac{1+\beta\cos\Theta}{1-\beta\cos\Theta}\right)^{\alpha+2}
\simeq
\left(\frac{4}{\Gamma^{-2}+\Theta^2}\right)^{\alpha+2}\,,
\end{equation}
where in the final expression the empirically well-justified
approximations $\Gamma\gg1$ and $\Theta\ll1$ were made.  Note that
this is a function of distance from the jet and in general will not be
constant near the black hole. 

For large Lorentz factors, and thus at large distances, $R_B$, is a
function of $\Theta$ alone.  The observed brightness ratio for the
putative counter jet is $\sim15$, and is roughly consistent with
the inclinations of $30^\circ$--$45^\circ$ obtained by radio proper
motion studies \citep{Reid_etal:89,Ly-Walk-Juno:07}.  This is, however,
much larger than the $20^\circ$ upper limit implied by apparent
motions of $\sim6c$ found by optical proper motion studies
\citep{Bire-Spar-Macc:99}.  There are many possible
explanations for this discrepancy, including asymmetric jets,
anisotropic emission, significant toroidal velocity components,
significant jet structure and variable spectral indexes, as well as
the misidentification of the counter jet.

Near the black hole, within the jet acceleration region $\Gamma$ can
be considerably smaller than the asymptotic value at large distances.
This is consistent with radio observations of moving features,
finding proper motions up to $2c$ within $2\,\mas$ from the core
\citep{Walk-Ly-Juno-Hard:08}, in contrast to the fast moving optical
features 500 times further away.  When $\Gamma\lesssim\Theta^{-1}$,
$R_B$ becomes a rapidly varying function of distance from the core,
providing an independent means to estimate the acceleration of the bulk
flow within the jet as a function of radius.

Utilizing the counter jet to study the jet acceleration and
orientation is necessarily predicated upon an accurate determination
of the location of the central supermassive black hole for two
reasons.  First, robustly identifying a given image feature as being
due to the counter jet is done most naturally by relating its position
to that of the black hole and jet.  Second, since the jet brightness
evolves as it moves away from the launching region, quantitatively
comparing the jet--counter jet brightnesses even for the idealized,
symmetric case, requires accurate distances from the central black
hole.  Positioning the supermassive black hole within the core via EHT
observations would remove any ambiguity regarding the black hole
location, and would undergird the interpretation of the putative
counter jet.

\subsubsection{Unambiguous Measurement of the Jet Opening Angle}
AGN jets are known to have remarkably small opening angles on large
scales.  This is set by both, the size of the jet launching region and
the rapidity with which the jet is collimated.  For M87, $7\,\mm$
images exhibit a radio core and a jet containing two limb-brightened
features, with opening angle $15^\circ$, commonly associated with the
opening angle of the jet itself 
\citep{Juno-Bire-Livi:99,Ly-Walk-Wrob:04,Kric_etal:06,Kova-List-Homa-Kell:07,Ly-Walk-Juno:07,Walk-Ly-Juno-Hard:08}.
However, these limb-brightened features do not appear to converge upon
the $7\,\mm$ core, intersecting roughly $2\,\mas$ east of it
\citep{Ly-Walk-Juno:07}.  One is then driven to one of three possible
conclusions, depending upon the location of the black hole
and the breadth of the jet launching region.

First, the $7\,\mm$ core does not indicate the position of the central
black hole, and is instead a feature within the inner jet, e.g., as
suggested by \citet{{Mars_etal:08}}.  In this case, it isn't clear why
the region surrounding the black hole doesn't radiate.  Nevertheless,
the $7\,\mm$ jet morphology would be broadly consistent with jet
formation theory.

Second, the $7\,\mm$ core does indicate the position of the central black
hole, and the jet rapidly collimates, begining from exceedingly large
opening angles ($\gtrsim60^\circ$) and reaching its asymptotic conical
structure within $\sim0.3\mas\simeq10^2 GM/c^2$ of the black
hole \citep{Juno-Bire-Livi:99}.  When the inclination exceeds the
largest opening angle, projection of the jet on the plane of the sky
significantly enlarges 
the apparent opening angle further, though this is insufficient to
explain the rapid evolution in the opening angle by itself.  In this
case, M87 presents a serious challenge to many jet formation theories,
and jet formation simulations in particular.

Third, the $7\,\mm$ core does indicate the position of the central black
hole, and the $7\,\mm$ jet seen on $\mas$ scales is associated with a
different structure than the optical jet seen at large distances.
This would naturally account for the systematically lower velocities
observed in the radio images.  In this case the transrelativistic
unbound wind seen in many jet formation simulations is a natural
candidate for the structure responsible for the $7\,\mm$ jet feature.

Locating the black hole with the EHT, and translating that location to
the longer-wavelength radio images, conclusively addresses the
distinction between the first two: either the black hole is located in
the radio core or not.  Furthermore, since the EHT promises to resolve
the jet launching region, it will be capable of distinguishing between
the compact ultra-relativistic jets and broad trans-relativistic
winds \citep{Brod-Loeb:09}.  By registering the mm-VLBI image to the
longer-wavelength VLBA images, the relationship between the anticipated
components of the outflow and the larger-scale features of the radio
images may be determined.

\subsubsection{Identifying the Source of Jet Variability}
M87 is known to be variable on timescales of weeks,
generating bright knots which rapidly propagate away from the radio
core.  The apparent speeds of these knots vary considerably, ranging
between $0.2c$ to $2.5c$
\citep{Ly-Walk-Juno:07,Kova-List-Homa-Kell:07,Walk-Ly-Juno-Hard:08}.
In at least one case these have been accompanied by a TeV flare
\citep{Acci_etal:09,Wagn_etal:09}, implying that the source of the
variability lies close to the black hole.  However, where the
variability originates remains to be determined.  Possible sources
include variations in the accretion rate, turbulence within the
accretion flow, and instabilities within the jet itself
\citep[see, e.g.,][]{Gian-Uzde-Bege:10}.
Monitoring M87 with coincident phase-referenced EHT and VLBA
observations allows variable features to be followed from horizon to
parsec scales.  Those arising near the black hole will be clearly
identifiable by their proximity to the silhouetted horizon.

\subsection{Possible Phase References}
\begin{figure*}
\begin{center}
\includegraphics[width=\textwidth]{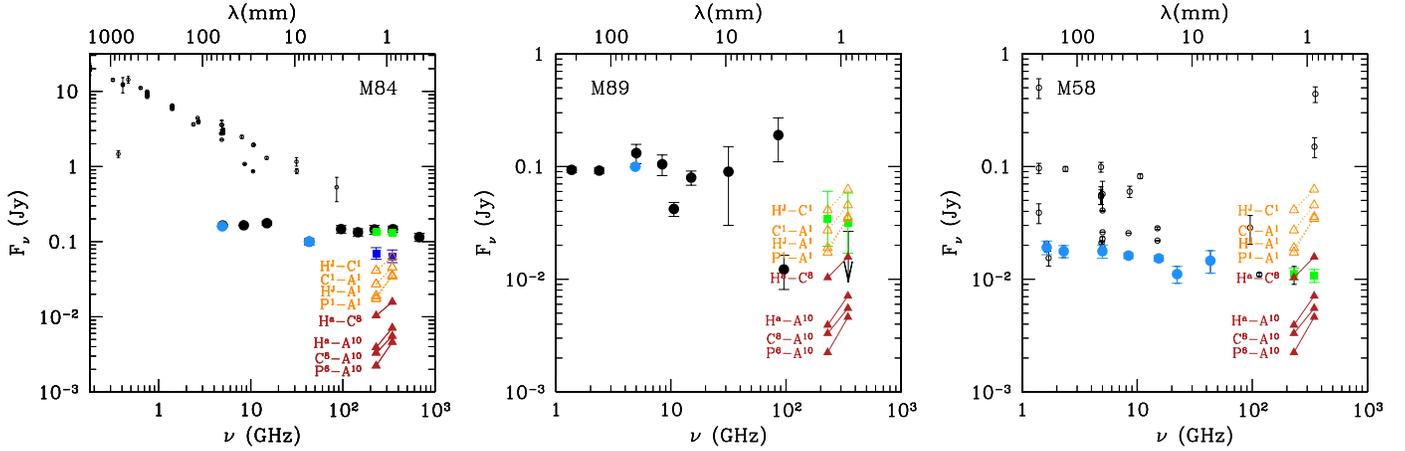}
\end{center}
\caption{SEDs of bright radio sources within $2^\circ$ of M87,
  obtained from the literature.
  Left: SED of M84.  The flux densities shown by the large filled circles
  correspond to observations with beam ${\rm FWHMs}$ smaller than
  $35''$.   Those shown in blue are VLBI observations, with typical
  beam sizes of $\mas$.  The $230\,GHz$ and $345\,\GHz$ fluxes implied
  by all of these points and the VLBI points alone are shown by the
  green and blue squares (respectively).  The small circles show all
  of the radio flux densities in the NED and VizieR databases associated
  with beams larger than $35''$, and are presumably indicative of the
  galactic non-thermal emission.
  Middle: SED of M89.  Large light blue circles show VLBI observations,
  while large black circles show all other radio observations in the NED and
  VizieR databases.  The green squares correspond to the $230\,\GHz$
  and $345\,\GHz$ flux densities implied by all observations.  For
  comparison, an upper limit at $347\,\GHz$ is shown by the arrow.
  Right: SED of M58.  Large light blue circles show VLBI
  observations, while small open circles show all other radio observations
  in the NED and VizieR databases.  The green squares are the
  $230\,\GHz$ and $345\,\GHz$ flux densities implied by the VLBI
  measurements.
  In all panels, for reference the sensitivities of baselines
  consisting of stations with multiple telescopes at both ends are
  shown at $230\,\GHz$ and $345\,\GHz$, by the open-orange and
  filled-red triangles (see the text for details).} \label{fig:M87SED}
\end{figure*}

Located well out of the Galactic plane, M87 does not suffer the
interstellar electron scatter-broadening and extinction associated
with studies of the Galactic center.  This has facilitated the
identification of a number of compact sources within it's vicinity.
However, few of these are radio sources, and fewer still have been
studied at VLBI resolutions.  As a consequence, the primary
uncertainty associated with the viability of potential mm-VLBI phase
references for M87 is their compactness.  Nevertheless, we have
identified a number of possible targets, which we list in order of
decreasing viability.

Located $1.52^\circ$ away from M87, M84 (NGC 4374) is one of
the few sources near M87 for which it is possible to find multiple VLBI
observations in the literature.  As with many of the phase references
we consider here, separating the contributions to the SED, shown in
Figure \ref{fig:M87SED}, associated with the surrounding galaxy and
the compact core is the primary difficulty.  In the case of M84
we do this via the resolution dependence of the spectral index.  The
SED obtained from observations found via NED\footnote{The NASA/IPAC Extragalactic Database (NED) is operated by the Jet Propulsion Laboratory, California Institute of Technology, under contract with the National Aeronautics and Space Administration.} and VizieR
\citep{VizieR:00} with large beams
\citep[${\rm FWHMs}>35''$, small open circles, ][]{Hees-Wade:64,Pilk-Scot:65,Paul-Wade-Hees:66,Gowe-Scot-Will:67,Eker:69,Kell-Paul-Will:69,Stul:71,Witz-Vero-Vero:71,Paul-Kell:72,Slee-Higg:73,Kell-Paul:73,Cond-Jaun:74,Hees-Conk:75,Sram:75,Will:75,Genz_etal:76,Vero:77,Paul_etal:78,Dres-Cond:78,Geld-Witz:81,Larg_etal:81,Kueh_etal:81,Benn_etal:86,Beck-Whit-Edwa:91,Greg-Cond:91,Whit-Beck:92,Slee:95,Doug_etal:96,Cond-Cott-Brod:02,Voll-Thie-Wiel:04,Cohe_etal:07} 
shows a characteristic steep power-law, almost certainly associated
with the galactic synchrotron component.  In contrast, the SED
associated with observations with small beams
\citep[${\rm FWHMs}<35''$, large filled circles, ][]{Leeu-Sans-Robs:00,Naga-Wils-Falc:01,Naga-Falc-Wils-Ulve:02,Leeu_etal:04,Ly-Walk-Wrob:04,Doi_etal:05,Naga-Falc-Wils:05},
and VLBI observations in particular
\citep[${\rm FWHMs}\sim\mas$, large filled blue circles, ][]{Ly-Walk-Wrob:04,Naga-Falc-Wils:05},
exhibit the flat spectrum indicative of AGN, extending from $5\,\GHz$
(the smallest frequency at which sufficiently small beam observations
have been performed) to $670\,\GHz$, above which dust emission begins
to dominate
\citep{Leeu-Sans-Robs:00,Leeu_etal:04,Doi_etal:05}.
Below $43\,\GHz$ the small-beam and VLBI flux densities are
similar, implying that in the radio, the small-beam observations are
dominated by the emission on sub-$\mas$ scales.  At far-infrared
wavelengths, \citet{Doi_etal:05} have excluded other potential sources of
contamination (galactic synchrotron, free-free and dust emission),
arguing that the dominant component is also due to the AGN itself.
As a result, the existing $0.85\,\mm$ ($350\,\GHz$) and $1.35\,\mm$
($221\,\GHz$) flux measurements, obtained via the JCMT with sub-$20''$
beams, provide a plausible estimate for the $230\,\GHz$ and
$345\,\GHz$ fluxes that will be observed by the EHT; finding values in
excess of $0.1\,\Jy$ (green-squares).  Nevertheless, we also
extrapolate flux estimates from the VLBI observations alone, finding
estimates on the order of $0.07\,\Jy$.  In both cases these are
considerably larger than the flux limits of the EHT-C at
$230\,\GHz$.  For the EHT-I, 
M84 is still almost certainly sufficiently bright to be used as
a phase reference at $230\,\GHz$, and is marginal at $345\,\GHz$.
As a result, M84 is likely to be a strong mm-VLBI phase
reference candidate.

M89 is somewhat closer to M87, at an angular separation of roughly
$1.19^\circ$.  The radio SED of M89 is shown in Figure
\ref{fig:M87SED}, in which all of the data points were taken from the
literature
\citep{Hees-Conk:75,Sram:75,Dres-Cond:78,Beck-Whit-Edwa:91,Greg-Cond:91,Naga-Wils-Falc:01,Cond-Cott-Brod:02,Filh_etal:04,Voll-Thie-Wiel:04,Doi_etal:05,Naga-Falc-Wils:05,Heal_etal:07,Brau_etal:07}
As with M84, there is a paucity of VLBI observations,
in this case only the one by \citet{Naga-Falc-Wils:05}, though the
resulting core flux density is similar to VLA observations at the same
frequency, differing by roughly 30\%
\citep[cf.][]{Beck-Whit-Edwa:91,Greg-Cond:91,Sram:75,Naga-Wils-Falc:01}.
This again suggests that the lower-resolution observations are
dominated by the compact core.  Unlike M84, there is a substantial
scatter in the high-frequency ($\sim90\,\GHz$) flux densities,
possibly due to resolution, an interpretation supported by the fact
that the relevant beam widths varied by an order of magnitude, with the
smaller beam measuring the smaller flux density
\citep[e.g.,][]{Hees-Conk:75,Doi_etal:05}.  This is unlikely to 
be responsible for the scatter at lower frequencies (near $10\,\GHz$),
where the higher-resolution observations produce higher flux density
estimates, though during a different epoch
\citep[e.g.,][]{Naga-Wils-Falc:01,Voll-Thie-Wiel:04}.  Therefore, to
infer the $230\,\GHz$ and $345\,\GHz$ flux densities we fit a single
power-law to all of the M89 radio data.  This results in values that
are consistent with the $347\,\GHz$ upper limit obtained by
\citet{Doi_etal:05}; sufficiently bright for the EHT-C, and only
marginal for the EHT-I.  Thus, we also consider this a strong
candidate mm-VLBI phase reference candidate, though clearly less so
than M84.

The final candidate we show in Figure \ref{fig:M87SED} is M58, again
with the radio SED obtained from the literature
\citep{Hees-Wade:64,Sram:75,Dres-Cond:78,Greg-Cond:91,Beck-Whit-Edwa:91,Whit-Beck-Helf-Greg:97,Naga-Wils-Falc:01,Ulve-Ho:01,Cond-Cott-Brod:02,Voll-Thie-Wiel:04,Ande-Ulve-Ho:04,Naga-Falc-Wils:05,Dale_etal:05,Doi_etal:05,Gall_etal:06,Krip_etal:07}.
Separated by $1.78^\circ$ from M87, M58 is the further than M84 and
M89, though still well within the isoplanatic angles at the
relevant EHT stations on good nights.  Unlike M84 and M89, sufficient
VLBI observations exist for M58 to completely characterize its radio
spectrum, from $1.7\,\GHz$ through $43\,\GHz$
\citep{Ande-Ulve-Ho:04,Naga-Falc-Wils:05,Krip_etal:07}.  These allow a
reasonably precise estimate of the mm-VLBI flux densities, giving
values comparable to the sensitivity limits of the EHT-C at
$230\,\GHz$, and well below them along the Hawaii--CARMA baseline at
$345\,\GHz$.  As a consequence, this is a marginal phase reference
candidate.

There are a number of additional bright sources near M87, though none
are well studied at sufficiently high resolutions to make definitive
statements regarding their millimeter fluxes on sub-$\mas$ scales.
Within $1^\circ$, the most promising candidates are J1229+114 (PKS
1227+119, VPCX 27) and J1230+114 (VPCX 28).  Extrapolating from the
existing radio data
\citep{Shim-Bolt-Wall:75,Larg_etal:81,Wrig-Otru:90,Beck-Whit-Edwa:91,Whit-Beck:92,Ledl-Owen:95,Greg-Scot-Doug-Cond:96,Whit-Beck-Helf-Greg:97,Laur_etal:97,Cond_etal:98,Reic_etal:00,Cohe_etal:07}
gives estimates for the millimeter fluxes
comparable to, or in excess of, the detection limits of the EHT-C,
though in most cases comparable to the limits for the EHT-I.  However,
these conclusions are contingent upon a substantial fraction of the
radio flux arising from sub-$\mas$ scales.
There is some evidence for the radio emission in J1229+114 being
primarily associated with the AGN, as opposed to  galactic
synchrotron; J1229+114 has been classified as an FRI based upon the
$1.4\,\GHz$ morphology as measured at 5'' resolutions by the FIRST
survey, performed with the Very Large Array
\citep{Whit-Beck-Helf-Greg:97,Gend-Wall:08}.  However, at 5'' it is
already apparent that the source is extended, with the unresolved core
constituting approximately 10\% of the overall flux
\citep{Whit-Beck-Helf-Greg:97}.
Similarly, \citet{Fles:10} assigns
a 41\% probability that J1230+114 is associated with a quasar based upon
radio--X-ray associations.  Furthermore, it is the only radio source
within 5' of MC2 1227+120, classified as a quasar in
\citet{Vero-Vero:83}, despite the lack of this source in the Molongo
catalogs \citep[cf. ][]{Larg_etal:81}.
Nevertheless, since neither object has been observed with
resolutions exceeding 5'', whether or not J1229+114 and J1230+114 are
suitable phase references remains highly uncertain.

Finally, many of the science objectives described in Section
\ref{sec:M87:SO} require only registering the mm-VLBI and cm-VLBI
reference frames.  As a consequence, many of these can be achieved
with astrometric accuracies of $\sim0.1\,\mas$.  Furthermore, due to the long
dynamical timescale of M87, it is possible to combine observations
over many nights to reduce the astrometric error.  However, as shown
in Appendix \ref{app:MACL}, the constraint upon the proximity of the phase
reference is dependent upon the astrometric precision desired and
the length of the observation.  Specifically, Equation \ref{eq:sigd}
implies that $0.1\,\mas$ precisions may be achieved at all relevant
EHT stations over the course of a single night, corresponding to
hundreds of independent $10\,\s$ measurements, for $\delta \lesssim
20^\circ$.  While this is sufficiently large that it violates some of
the assumptions made in deriving Equation \ref{eq:sigd}, it is
nevertheless clear that angular separations much larger than a few
degrees are permissible for this purpose.  Thus, we include a fifth
possible reference, the well-studied bright quasar 3C 273, located
$10.3^\circ$ away from M87.
A sufficient number of VLBI flux measurements exist for 3C 273 to
produce an SED from these alone
\citep{Kric_etal:97,Lons-Doel-Phil:98,Kell_etal:04,Zava-Tayl:04,Hori_etal:04,List-Homa:05,Kova_etal:05,Cara-List:08,Lee_etal:08}.
They include the observation by \citet{Kric_etal:97} at $215\,\GHz$
which found a flux density of $9.2\pm0.6\,\Jy$.  As a result, there is
no question that 3C 273 is bright enough at millimeter and
sub-millimeter wavelengths to be used as an astrometric reference.

\section{Conclusions}  \label{sec:C}

The EHT promises to usher in an era of $\muas$ black hole astronomy.
This takes the form first of $\sim20\,\muas$-resolution images,
allowing careful study of the morphology and dynamics of accretion
flows and jet formation on sub-horizon scales for a handful of
supermassive black holes, including Sgr A* and M87.  Because these
images resolve the horizon, they unambiguously locate the black hole
via its silhouette against nearby emission.  However, it also allows
high-precision astrometric observations, relating the positions of Sgr
A* and M87 to nearby, bright, compact sources.  This is only possible
because many of the EHT VLBI stations are arrays, allowing coincident
observations of the source and a reference without having to repoint
the antennas on the atmospheric coherence time, roughly $10\,s$.  On
a typical night, atmospheric turbulence limits the precision to
$4\,\muas$ over $10\,\min$ and $1\,\muas$ over many hours.

Necessary for $\muas$-precision astrometry with the EHT is nearby,
bright point sources.  This is complicated by both the short
wavelength (few objects are bright in the mm) and the very high
angular resolution of the EHT (many objects are resolved out).
Nevertheless, with the anticipated $4\,\GHz$ bandwidths a variety of
sufficiently bright possible phase references exist near Sgr A* and
M87.  While in most cases we cannot determine the mm-size of these
objects, in all cases they have not been resolved out at $7\,\mm$.
Note that we have concerned ourselves solely with objects that can be
detected individually on timescales shorter than the atmospheric
coherence time, and thus the sensitivity/brightness limits can be
relaxed if these objects are phase referenced by the primary source
(e.g., Sgr A* or M87).  

The kinds of science that $\muas$-precision astrometry enables depends
critically upon the object under consideration.  Many of the processes
that potentially affect the position of the central, supermassive
black hole depend upon its mass, while the questions of interest
depend upon its environment.  In all cases, however, the unique
ability of mm-VLBI to conclusively locate the position of the
supermassive black hole is used.  

Sgr A* is constantly buffeted by objects within the evolved stellar
cusp.  Even with extremely pessimistic assumptions regarding the
content and structure of the cusp, over year timescales, this produces
$\muas$ displacements in Sgr A*'s position.  While the presence of
these displacements is robust, the magnitude is sensitive to the
properties and density of the objects within the cusp, and thus
provides a means to probe these.  Detection of the expected
stellar-mass black hole cusp would have far reaching implications for
the event rates of existing and future gravitational wave searches.

Sgr A* is also potentially perturbed by massive binary companions.
Frequently invoked to explain the presence of young, massive stars
within $10^2\,\AU$ of Sgr A*, an intermediate black hole orbiting Sgr
A* has an infall timescale (dominated by gravitational radiation) that
is long in comparison to the lifetimes of the young stars.  If such a
binary exists it will induce periodic displacements in Sgr A*'s
position, the size of which depend upon the object's mass and orbital
parameters.  While some portion of the available parameter space has
previously been excluded by $7\,\mm$ astrometric studies, the
$\muas$-precision astrometry made possible with the EHT is capable of
eliminating the vast majority of the remaining parameters with
sub-decade observations.

Efforts to image Sgr A* must necessarily contend with its sub-hour,
large-amplitude variability.  This formally precludes Earth aperture
synthesis, which assumes that the source is fixed throughout the
night and is not amenable to self-calibration techniques, which
smear dynamic images in an ill-defined fashion.  Phase reference
observations provide a way in which to faithfully relate visibilities
during variable episodes, providing a time-averaged image of the Sgr
A*.  In the absence of an a priori accurate model for Sgr A*'s flares,
this makes it possible to identify the qualitative features of the
relevant processes.

Due to its much larger mass, what can be probed via astrometry for M87
is very different than what is accessible for Sgr A*.  In M87's case,
perturbations induced by the stellar/remnant cusp are negligible, as
are displacements due to potential binary companions.  Instead, the
utility of astrometry with the EHT is the ability to relate the
mm-VLBI image, in which the black hole is clearly identifiable due to
its silhouette against nearby emission, and the large-scale structures
in longer-wavelength images.  This provides a way in which to
conclusively locate the black hole, eliminating ambiguity in the
identification of a putative counter jet, inclination, opening angle,
nature of the radio core, and relationship between the black hole and
the radio jet.  Moreover, during dynamical periods, with the EHT it
is possible to follow the creation and propagation of radio knots from
sub-horizon to parsec scales, shedding light upon the source of the
observed variability in M87 specifically, and AGN more generally.

\acknowledgments
We thank Irwin Shapiro, Sheperd Doeleman and Vincent Fish
for enlightening discussions and suggestions.
This research has made use of the NASA/IPAC Extragalactic Database
(NED) which is operated by the Jet Propulsion Laboratory, California
Institute of Technology, under contract with the National Aeronautics
and Space Administration.
This work was supported in part by NSF grant AST-0907890 and NASA grants
NNX08AL43G and NNA09DB30A.

\appendix
\section{Millimeter Isoplanatic Angle} \label{app:MACL}
Understanding atmospheric propagation effects is critical to
high-fidelity image reconstruction at high frequencies.  As a result,
considerable effort has been expended theoretically and empirically
characterizing the phase fluctuations associated with spatially
varying indices of refraction in the turbulent atmosphere.  A general
treatment can be found in \citet{Thom-Mora-Swen:01}; here we collect
some results that are of interest to mm-VLBI astrometry.

We wish to estimate the limits imposed by atmospheric propagation
effects upon the accuracy with which the relative locations of nearby
sources may be determined at $230\,\GHz$ ($1.3\,\mm$) and $345\,\GHz$
($0.87\,\mm$).  To do this we will make a number of simplifying
approximations and convenient assumptions regarding the structure of
the intervening atmosphere and the sources of interest.  We begin by
defining a model for the atmospheric fluctuations, and the associated
phase errors along nearby lines of sight.  After doing so, we will
specify an idealized astrometric experiment, and compute the
associated errors induced by the forgoing atmospheric model.

\begin{deluxetable*}{llcccccl}\tabletypesize{\small}
\tablecaption{Phase Error Structure Function Parameters and
  Isoplanatic Angles at Potential EHT Stations\label{tab:sigd0}}
\tablehead{
\colhead{Site} &
\colhead{Telescopes} &
\colhead{Alt.} &
\colhead{$\sigma_{d0}$\tablenotemark{a}} &
\colhead{$\beta$\tablenotemark{b}} &
\colhead{$\delta_0^{230\,\GHz}$} &
\colhead{$\delta_0^{345\,\GHz}$} &
\colhead{Reference}\\
&
&
\colhead{(m)} &
\colhead{(mm)} &
&
&
&
}
\startdata
Mauna Kea\tablenotemark{c}       & JCMT, SMA, CSO      & 4070 & 0.4--2.7 & 0.75     & $0.6^\circ$--$7.7^\circ$  & $0.4^\circ$--$4.5^\circ$  & \citet{Mass:94}\\
Chile\tablenotemark{c}           & ASTE, APEX, ALMA    & 5000 & 0.3--1.5 & --       & $1.6^\circ$--$11^\circ$  & $1.0^\circ$--$6.6^\circ$  & \citet{NRAO:98}\\
Plateau de Bure\tablenotemark{c} & IRAM Interferometer & 2552 & 0.3--0.7 & 1.1--1.9 & $5.3^\circ$--$9.6^\circ$  & $4.3^\circ$--$6.7^\circ$  & \citet{Olmi-Down:92}\\
Cedar Flat\tablenotemark{d}      & CARMA               & 2196 & 0.8--3.8 & --       & $0.5^\circ$--$3.3^\circ$  & $0.3^\circ$--$2.0^\circ$  & \citet{Wood_etal:04}
\enddata
\tablecomments{Adapted from \citet{Thom-Mora-Swen:01}, Table 13.3.}
\tablenotetext{a}{Defined at a baseline of $1\,\km$.}
\tablenotetext{b}{Where no value was given, $\beta=5/6$, consistent with Kolmogorov turbulence, was assumed.}
\tablenotetext{c}{Ranges indicate good/bad observing conditions (e.g., nighttime/daytime).}
\tablenotetext{d}{Ranges indicate before and after water vapor radiometry has been employed.}
\end{deluxetable*}

The atmospheric model is defined by fluctuations on top of some mean
structure.  We make the common assumption of a plane parallel
atmosphere, imparting an average phase delay.  Superimposed upon this
are gaussian phase fluctuations, assumed to be due to a thin layer
within the troposphere (fixed at an altitude of roughly
$L\simeq1\,\km$).  The resulting phase perturbations are then
characterized by the mean value and the spatial correlation function,
commonly referred to as the structure function.  These are defined by
their zenith values:
\begin{equation}
\begin{gathered}
\Phi_Z \equiv \langle \phi_Z(\bx,t) \rangle\,,\\
\mathcal{D}_\phi(\left|\bd\right|) \equiv \langle \left[ \phi_Z(\bx+\bd,t) -  \phi_Z(\bx,t) \right]^2 \rangle\,,\\
\Sigma^2_\phi\equiv\langle \phi_Z^2(\bx,t)\rangle - \Phi_Z^2
= \frac{1}{2} \mathcal{D}_\phi(\infty)\,,
\end{gathered}
\end{equation}
where we have given the auto-correlation function a unique symbol
despite being degenerate with the $\mathcal{D}_\phi(d)$, and $\bd$
will correspond to interferometer baselines.  Each of
these quantities is a property of the particular site and may be
empirically determined.  For example, $\Phi_Z$ may be obtained from
various atmospheric models
\citep[e.g.,][]{Niel:96,Boeh-Neil-Treg-Schu:06,Boeh-Werl-Schu:06},
with typical values of $(600\,\cm/\lambda)\,\rad$ (though this varies
by roughly a factor of two between sea level and $5000\,\m$, the
altitude of the ALMA and Hawaii sites).  At small spatial separations
the structure function has been found to be well fit by a power law,
$\mathcal{D}_\phi(d)\simeq \left(2\pi\sigma_{d0}/\lambda\right)^2
(d/1\,\km)^{2\beta}\,\rad$ with values for $\sigma_{d0}$ and $\beta$
given in Table \ref{tab:sigd0} for the relevant locations.  At large
$d$, $\mathcal{D}_\phi(d)$ asymptotes to a fixed value,
$2\Sigma_\phi^2$.  Typical values for $\Sigma_\phi$ are
$0.01\Phi_Z\simeq(6\,\cm/\lambda)\,\rad$, is dominated by large-scale
variations in the wet component and varies substantially on hourly
timescales.

The phase perturbation along a general oblique line of sight, defined
by a position on the sky $\balpha$, associated with zenith angle 
$z_{\small\bx,\balpha}$ at location $\bx$, in a plane-parallel
geometry is given by
\begin{equation}
\phi_a(\bx,\balpha,t) = \phi_Z(\bx+L\tan z_{\small\bx,\balpha},t) \sec z_{\small\bx,\balpha}\,.
\end{equation}
From this we can construct an angular structure function:
\begin{equation}
\begin{aligned}
\mathcal{D}_\phi(\balpha,\bdelta)
&\equiv
\langle \left[ \phi_a(\bx,\balpha+\bdelta,t) - \phi_a(\bx,\balpha,t) \right]^2 \rangle\\
&\qquad\qquad\qquad
- \Phi_Z^2\sec^2 z\, \tan^2 z\,\,{\delta z}^2\\
&\simeq
\mathcal{D}_\phi(d) \sec^2 z
\left[ 1 + \tan z \,\delta z + \frac{1+\sin^2 z}{2\cos z} {\delta z}^2 \right]\\
&\qquad\qquad\quad
+\Sigma_\phi^2 \sec^2 z \,\tan^2 z \,\,{\delta z}^2
+ \mathcal{O}\left({\delta z}^3\right)\,,
\end{aligned}
\label{eq:asf}
\end{equation}
where we have suppressed the subscripts on $z$ and $\delta z$, which
should be understood to be uniquely defined in terms of $\bx$,
$\balpha$, and $\bdelta$,
\begin{equation}
\begin{aligned}
d
&\equiv
L\sqrt{
  \left[\tan(z+\delta z)-\tan z\right]^2
  +
  \left|\bdelta\right|^2
  -{\delta z}^2
}\\
&\simeq
L \sqrt{
  \left(\sec^4z - 1\right){\delta z}^2
  +
  \left|\bdelta\right|^2
}\,,
\end{aligned}
\end{equation}
and we have made use of the Taylor expansion  
of $\sec(z+\delta z)$ to 2nd order:
\begin{equation}
\sec(z+\delta z)
=
\sec z
\left[
1
+ 
\tan z \,\delta z
+
\frac{1+\sin^2 z}{2\cos z} {\delta z}^2
+
\mathcal{O}\left({\delta z}^3\right)
\right]\,.
\end{equation}
Unlike $\mathcal{D}_\phi(d)$, which is completely defined by the
small-scale structure of the troposphere,
$\mathcal{D}_\phi(\balpha,\bdelta)$ has a contribution due to
large-scale atmospheric variations as a result of the different path
lengths along the two lines of sight, resulting from the slightly
different zenith angles.  Which term dominates in Equation
(\ref{eq:asf}) depends upon the zenith angle, how strongly nearby
atmospheric patches are correlated and the magnitude of the
large-scale atmospheric fluctuations about the average phase delay.

To estimate the limits upon the accuracy this implies we consider a
simplified astrometric experiment.  Specifically, we will assume that
the two sources of interest are monochromatic point sources separated
by more than a beam width, though extension to more complicated source
structures is straightforward.  Furthermore, we focus only upon
atmospheric phase errors, and explicitly neglect all other
contributions.  Finally, we will assume the experiment consists of two
sets of antenna pairs, one pair located at $\bx=\bmath{0}$ and the
other pair at some position $\bx=\bmath{u}$, sufficiently far away
such that the atmospheric errors are completely uncorrelated.  With
these assumptions, the voltages induced by the radio waves emitted by
the two point sources, located at positions $\alpha$ and
$\alpha+\delta$, at a pair of neighboring antennas at some position
$x$, each viewing one of the sources, are 
\begin{equation}
\begin{aligned}
V_A(\bx,t) &=
E_I \exp\left\{-i\left[\omega t + \frac{2\pi\bx}{\lambda}\cdot\balpha + \phi_a(\bx,\balpha,t)\right]\right\}\\
V_B(x,t) &=
E_{II} \exp\left\{-i\left[\omega t + \frac{2\pi\bx}{\lambda}\cdot(\balpha+\bdelta) + \phi_a(\bx,\balpha+\bdelta,t)\right]\right\}\,,
\end{aligned}
\end{equation}
where $E_{I,II}$ are two complex constants,
$\phi_a(\bx,\balpha,t)$ are the random phase perturbations imparted by
propagation through the atmosphere defined above, and the remaining
term is the geometric phase delay that we seek to measure.
Cross-correlating the antenna voltages along the long baselines
produces correlation peaks for each source at time delays of
\begin{equation}
\begin{aligned}
\tau_A &= \frac{2\pi\bu}{\lambda}\cdot\balpha + \phi_{a,\small\bu}(\balpha,t)\\
\tau_B &= \frac{2\pi\bu}{\lambda}\cdot(\balpha+\bdelta)  + \phi_{a,\small\bu}(\balpha+\bdelta,t)\,,
\end{aligned}
\end{equation}
where $\phi_{a,\small\bu}(\balpha,t) \equiv \phi_a(\bu,\balpha,t)-\phi_a(\bmath{0},\balpha,t)$,
in which we have assumed that the atmospheric phase delay varies slowly in
comparison to the timescales over which the visibilities are
measured.  From these the relative positions of the point sources may be
extracted in the normal way, obtaining an estimate of the projection
of the angular distance between two sources along $\bu$ of
\begin{equation}
\begin{aligned}
\bhu\cdot\bdelta_{\rm obs}
&\equiv
\frac{\lambda}{2\pi u}\left(\tau_B-\tau_A\right)
-
\Delta
\\
&=
\bhu\cdot\bdelta +
\frac{\lambda}{2\pi u}
\left[
  \phi_{a,\small\bu}(\balpha+\bdelta,t)-\phi_{a,\small\bu}(\balpha,t)
  \right]
-
\Delta
\\
&=
\bhu\cdot\bdelta +
\frac{\lambda}{2\pi u}
\left\{
\left[\phi_{a}(\bu,\balpha+\bdelta,t)-\phi_{a}(\bu,\balpha,t)\right]\right.\\
&\qquad\qquad\left.-
\left[\phi_{a}(\bmath{0},\balpha+\bdelta,t)-\phi_{a}(\bmath{0},\balpha,t)\right]
\right\}
-
\Delta
\,,\\
\end{aligned}
\end{equation}
where
$\Delta\equiv \left.\Phi_Z\sec z\,\tan z\,\delta z\right|_{\small\bmath{u}}-\left.\Phi_Z\sec z\,\tan z\,\delta z\right|_{\small\bmath{0}}$
is the angular shift due to the differential phase delays at the
two stations associated with the mean atmospheric model.  The
$\bhu\cdot\bdelta_{\rm obs}$ is a gaussian random variable centered
upon the true value and with a variance due to the atmospheric phase
fluctuations of
\begin{equation}
\sigma_{\small\bdelta}^2
=
\left(\frac{\lambda}{2\pi u}\right)^2
\left[
  \left.\mathcal{D}_\phi(\balpha,\bdelta)\right|_{\small\bmath{0}}
  +
  \left.\mathcal{D}_\phi(\balpha,\bdelta)\right|_{\small\bmath{u}}
\right]\,.
\end{equation}
This has contributions from both, the small-scale, rapidly varying
structures that limit the atmospheric coherence time (the
$\mathcal{D}_\phi(d)$ terms in Equation (\ref{eq:asf})), and
large-scale, slowly evolving fluctuations (responsible for the
$\Sigma_\phi$ terms in Equation (\ref{eq:asf})).  Because $\beta$ is
typically close to unity, which dominates the astrometric uncertainty
is only weakly dependent upon the angular separation of the two
sources.  In the remainder of this section we discuss the two kinds
of terms separately.

The error term associated with the rapidly varying, small-scale
structures due to a single station is given by
\begin{equation}
\sigma_s \lesssim \frac{\lambda}{u} \frac{\sigma_{d0}}{\lambda} \sec z
\left(\frac{L}{1\,\km}\,\sec^2z\,\,\delta z\right)^{\beta}\,,
\end{equation}
where we have chosen $\delta z=\left|\bdelta\right|$, resulting in the maximum
$\sigma_s$ for a given angular separation.  For a values of $\sigma_{d0}$,
$\beta$, and $z$ typical
of the stations likely to be employed by the EHT, at $1.3\,\mm$,
$\sigma_s\lesssim 0.5(\lambda/u) (L/1\,\km)^\beta (\delta z/1^\circ)^\beta$.

Since $\sigma_s$ is dominated by contributions from the most rapidly
varying structures in the atmosphere, observations separated in time
by more than the atmospheric coherence time probe independent
realizations of the atmospheric turbulence.  As a consequence, during
a few hour observing period it is possible to reduce $\sigma_s$ by an
order of magnitude.  Thus, the limit upon $\delta$ imposed by a given
astrometric tolerance, $\sigma_A$, and $N$ independent observations is
\begin{equation}
\delta \lesssim \delta_0
\equiv
57^\circ \,\cos^2 z \left(\frac{L}{1\,\km}\right)^{-1}
\left(\sqrt{N} \cos z \,\frac{u\sigma_A}{\sigma_{d0}}\right)^{1/\beta}
\end{equation}
Limits for $N=1$ and $\sigma_A=\lambda/u$ are listed in Table
\ref{tab:sigd0} for $1.3\,\mm$ and $0.87\,\mm$ for $z=70^\circ$,
typical of the zenith angle for Sgr A* as seen from North American
sites (the most pessimistic scenario we consider).  In terms of these,
we have
\begin{equation}
\sigma_s \simeq \frac{2.9\cos z}{\sqrt{N}} \left(0.12\sec^2 z\,\frac{\delta}{\delta_0}\right)^\beta \frac{\lambda}{u}\,.
\label{eq:sigd}
\end{equation}

The error associated with the large-scale, slowly evolving component
at a single station is given by
\begin{equation}
\sigma_l \simeq \frac{\lambda}{u} \frac{\Sigma_\phi}{2\pi} \sec z\,\tan z\,\,\delta z\,.
\end{equation}
For Sgr A* viewed from North America, typically
$\sigma_l\lesssim (\lambda/u) (\delta z/1^\circ)$, can be 
substantially smaller at high-altitude sites, and is
a strong function of zenith angle.

Unlike the small-scale component, the long timescale over which the
large-scale structures vary means that only a handful of independent
realizations will be encountered on a given night.  Thus,
substantially reducing $\sigma_l$ by integrating over long times requires
observations spanning many days at least.  However, as we discuss in
the main text, the large spatial and temporal scales of these fluctuations
make possible a number of schemes to measure the additional path delay
directly and significantly reduce the unmodeled $\Sigma_\phi$.  For
this reason the uncertainty in the angular separation due to unmodeled
slowly varying and rapidly varying components,
\begin{equation}
\sigma_\delta^2
=
\left.\sigma_s^2\right|_{\small\bmath{0}}
+
\left.\sigma_l^2\right|_{\small\bmath{0}}
+
\left.\sigma_s^2\right|_{\small\bmath{u}}
+
\left.\sigma_l^2\right|_{\small\bmath{u}}\,,
\end{equation}
are likely to be similar.

\section{Multi-Reference Astrometric Calibration} \label{app:MRAC}
Here we describe explicitly a scheme to measure and remove the
contribution to the astrometric uncertainty arising from slowly
varying, large-scale atmospheric fluctuations (i.e., those responsible
for the $\sigma_l$ terms in the preceding appendix).  The method
utilizes multiple nearby reference sources to measure the
slowly-varying phase fluctuations, which may then be explicitly
accounted for in the astrometric measurement of interest.  The
viability of the scheme is predicated upon the following
assumptions:
\begin{itemize}
\item The existence of a set of at least three reference sources near the
  target source with different zenith angles at each VLBI station.
  Note that the limitations upon the separation
  derived in Section \ref{app:MACL} may be substantially relaxed since
  both $\sigma_l$ and $\sigma_s$ are both roughly proportional to
  $\left|\bdelta\right|$, and thus the limit upon the fractional
  accuracy imposed by rapidly-varying, short timescale atmospheric
  fluctuations is only weakly dependent upon separation.  In practice
  $\left|\bdelta\right|$ is limited by the size of the region over
  which the large-scale phase perturbations are uniform, which is
  typically many degrees.
\item Over the portion of the sky of interest, the large-scale
  atmospheric phase delays at each site is well modeled by a
  single-parameter model (e.g., the $\sec z$ model we will employ
  here).
\item The reference sources do not evolve intrinsically or relative
  to each other.  More specifically, the angular separation between
  each of the reference sources is fixed.
\item The ``true'' angular separations of the reference sources can be
  measured with the desired fractional accuracy.  This may be
  accomplished, e.g., by careful atmospheric modeling and long-time
  integrations.
\end{itemize}

To illustrate the procedure, and estimate the resulting calibration
uncertainties we consider observations of three reference sources on a
single, static baseline.  We use the letters $A$, $B$, and $C$ to
indicate a particular source, and numbers $1$ and $2$ to indicate the locations
of individual mm-VLBI stations.  For short-hand we define
$\left.\Delta\sec z\right|_{B1}^{A1}=\sec z_{A1} - sec z_{B1}$ to be
the difference between the secant of the zenith angle of reference
sources $A$ and $B$ as seen from station $1$, with similar expressions
for all sources and both stations.  In terms of these, following the
previous appendix, from a single observation we have an estimate of
the separation of reference sources $A$ and $B$ of:
\begin{multline}
\bhu_{12}\cdot\bdelta_{AB,\rm obs}
=
\bhu_{12}\cdot\bdelta_{AB}
+
\Delta\delta_{s,AB}\\
+
\frac{\lambda}{2\pi u_{12}}\Delta\Phi_1 \left.\Delta\sec z\right|_{B1}^{A1}
-
\frac{\lambda}{2\pi u_{12}}\Delta\Phi_2 \left.\Delta\sec z\right|_{B2}^{A2}\,,
\end{multline}
where we have separated the contributions to the phase fluctuation
into modeled large-scale and the remaining unmodeled small-scale
fluctuations ($\Delta\delta_s$).  The $\Delta\Phi_{1,2}$ are the
unknown zenith phase delays implied at stations $1$ and $2$,
respectively, associated with the large-scale atmospheric
perturbations.  Since $\bdelta_{AB}$ is known, or may be measured by
comparing data sets taken over many days, we may invert this to
obtain an expression for the unknown $\Delta\Phi_{1,2}$:
\begin{equation}
\begin{aligned}
&\Delta\Phi_1 \left.\Delta\sec z\right|_{B1}^{A1}
-
\Delta\Phi_2 \left.\Delta\sec z\right|_{B2}^{A2}\\
&\qquad\qquad\qquad\quad
=
\frac{2\pi}{\lambda}\left[
\bu_{12}\cdot\left(\bdelta_{AB,\rm obs}-\bdelta_{AB}\right)
-
\Delta\delta_{s,AB}\right]\\
&\Delta\Phi_1 \left.\Delta\sec z\right|_{C1}^{B1}
-
\Delta\Phi_2 \left.\Delta\sec z\right|_{C2}^{B2}\\
&\qquad\qquad\qquad\quad
=
\frac{2\pi}{\lambda}\left[
\bu_{12}\cdot\left(\bdelta_{BC,\rm obs}-\bdelta_{BC}\right)
-
\Delta\delta_{s,BC}\right]\\
&\Delta\Phi_1 \left.\Delta\sec z\right|_{A1}^{C1}
-
\Delta\Phi_2 \left.\Delta\sec z\right|_{A2}^{C2}\\
&\qquad\qquad\qquad\quad
=
\frac{2\pi}{\lambda}\left[
\bu_{12}\cdot\left(\bdelta_{CA,\rm obs}-\bdelta_{CA}\right)
-
\Delta\delta_{s,CA}\right]\,,
\end{aligned}
\label{eq:spc1}
\end{equation}
where we have added analogous terms for the other source pairs.  In
principle, these are degenerate, providing only two linearly
independent equations for $\Delta\Phi_1$ and $\Delta\Phi_2$.  In
practice, since the $\Delta\delta_s$ terms are dominated by the
small-scale atmospheric structure, this degeneracy is weakly broken,
providing an external estimate of the magnitude of the
$\Delta\delta_s$ terms.  Considering only the first two, we obtain an
estimate for $\Delta\Phi_1$ of
\begin{multline}
\Delta\Phi_1
=\\
\frac{2\pi\bu_{12}}{\lambda}
\cdot
\frac{
\left.\Delta\sec z\right|_{B2}^{A2}
  \left(\bdelta_{AB,\rm obs}-\bdelta_{AB}\right)
-
\left.\Delta\sec z\right|_{C2}^{B2}
  \left(\bdelta_{BC,\rm obs}-\bdelta_{BC}\right)
}
{
\left.\Delta\sec z\right|_{B2}^{A2}
\left.\Delta\sec z\right|_{C1}^{B1}
-
\left.\Delta\sec z\right|_{B1}^{A1}
\left.\Delta\sec z\right|_{C2}^{B2}
}
\end{multline}
with a similar expression for $\Delta\Phi_2$ obtained by transposing
$1$ and $2$ everywhere.  Comparing this to Equation (\ref{eq:spc1}),
assuming that the errors in the $\bdelta_{AB,BC}$ may be neglected in
comparison to $\Delta\delta_s$ (the errors in the zenith angle
estimates result in higher-order corrections, and thus always may be
ignored), we find a variance in $\Delta\Phi_1$ induced by the
small-scale, rapid variability of
\begin{equation}
\sigma_{\Delta\Phi_1}^2
\simeq
\frac{
\left(\left.\Delta\sec z\right|_{B2}^{A2}\right)^2
\sigma_{s,AB}^2
+
\left(\left.\Delta\sec z\right|_{C2}^{B2}\right)^2
\sigma_{s,BC}^2
}
{
\left(
\left.\Delta\sec z\right|_{B2}^{A2}
\left.\Delta\sec z\right|_{C1}^{B1}
-
\left.\Delta\sec z\right|_{B1}^{A1}
\left.\Delta\sec z\right|_{C2}^{B2}
\right)^2
}\,,
\end{equation}
with a similar expression for $\sigma_{\Delta\Phi_2}^2$.  Note that if
all of the reference sources lie at similar zenith angles, and have
similar separations in zenith angle from each other, this reduces to
$\sigma_{\Delta\Phi_{1,2}}\sim\sigma_s$, and thus the fractional
precision of the large-scale phase corrections is comparable to that
associated with the small-scale, rapidly varying fluctuations over
hour-long integrations, roughly $1\,\muas$.  The accuracy will be
limited by how well the true separations of the reference sources can
be measured, presumably by averaging over long times, and the validity
of the assumptions underlying the scheme.

\section{Position Jitter Power Spectrum} \label{app:PJPS}
Supermassive black holes settle via dynamical friction into the center
of their galactic potential well on timescales short in comparison to
the Hubble time \citep{Chat-Hern-Loeb:02}.  Nevertheless, fluctuations
in the surrounding stellar/remnant\footnote{Henceforth, for
  simplicity we will refer to the ``stellar'' distribution,
  ``stellar'' mass, and
  ``stars'' though this should be understood to include evolved
  objects like stellar mass black holes as well.} distribution induce
random displacements of the central black hole similar to Brownian
motion.  Here we derive an estimate of the displacement power
spectrum.  For simplicity we assume that the surrounding distribution
of gravitating bodies is spherically symmetric, with number density
$n(r)$ and mass function $\phi(m,r)$, all orbits are circular, and
that dynamical collective N-body effects are unimportant, i.e., the
fluctuations about the mean density may be approximated by Poisson
statistics.  This last assumption may not be strictly justified in the
spherical case, though is in the more generic triaxial situation.

In the vicinity of the supermassive black hole, we may approximate the
black hole's motion by
\begin{equation}
\ddot{\bmath{x}} + \gamma \dot{\bmath{x}} + \omega_0^2 \bmath{x}
=
\bmath{a}(t)\,,
\end{equation}
where $\bmath{x}$ is the location of the black hole, $\bmath{a}$ is
the fluctuating acceleration induced by perturbations in the
surrounding stellar distribution, $\omega_0^2 \equiv \d^2\Phi/\d r^2$ is
the natural frequency of oscillation in the bottom of the galactic
potential well, and $\gamma$ is the damping associated with dynamical
friction.  We may estimate $\omega_0$ in terms of the stellar density
at the galactic center:
\begin{equation}
\omega_0^2 \simeq \frac{4\pi}{3} G \rho_0
\end{equation}
where $\rho_0$ is the average stellar density within the region of
influence.  Thus, $\omega_0$ is a proxy for the stellar mass
density within the core.  The dynamical damping rate is approximately
given by
\begin{equation}
\gamma \simeq 4\sqrt{2\pi} \ln\Lambda G^2 M \frac{\rho_0}{\sigma_0^3}
\end{equation}
where $\ln\Lambda$ is the Coulomb integral,  $M$ is the mass of the
central supermassive black hole, and $\sigma_0$ is the stellar
velocity dispersion in the stellar core \citep{Chat-Hern-Loeb:02}.
Typically, we will be interested in the case when
$\omega\gg\omega_0,\,\gamma$, though we have kept all terms in the
interest of generality.

Since the system is spherically symmetric, we will focus only upon the
displacement along the $\bmath{\hat{x}}$ direction.  In this case, the
power spectrum of these displacements is then given by
\begin{equation}
P_{x,\omega}
=
\langle \left|x_\omega\right|^2 \rangle
=
\frac{P_{a,\omega}}{ (\omega^2-\omega_0^2)^2 + \gamma^2\omega^2 }
\simeq
\omega^{-4} P_{a,\omega}
\,,
\label{eq:Pxw}
\end{equation}
where $P_{a,\omega}$ is the power spectrum of the fluctuating
accelerations.  Thus, computing the spectrum of displacements is
reduced to computing the spectrum of the fluctuating forces.

In \citet{Chat-Hern-Loeb:02} it was assumed that the fluctuations were
completely uncorrelated in time, i.e., that $P_{a,\omega}$ was
constant.  For virialized systems this may be derived directly via the
fluctuation dissipation theorem.  However, this is likely to be a poor
approximation in the case of interest, where we are interested
exclusively in short-timescale variations in the position of the
supermassive black hole.  Thus we approach the problem
phenomenologically: given a density profile we estimate the
$P_{a,\omega}$ associated with Poisson fluctuations in the number
density of stars.

We begin by considering a thin shell of width $\Delta r$ centered upon
a radius $r$.  Within this shell there are $N=4\pi r^2 n(r) \Delta r$
objects, with mass function $\phi(m,r)$, randomly distributed.  The
induced acceleration along the $\bmath{\hat{x}}$-axis is then 
\begin{equation}
a = -\sum_{i=1}^N \frac{G m_i x_i}{r^3}\,,
\end{equation}
from which it is obvious that $\langle a \rangle = 0$.  The variance
in the acceleration is
\begin{equation}
\begin{aligned}
\langle a^2 \rangle
&=
\langle \sum_{i,j=1}^N \frac{G^2 m_i m_j x_i x_j}{r^3} \rangle
=
\frac{G^2}{r^6} \sum_{i,j=1}^N \langle m_i m_j \rangle \langle x_i x_j\rangle \\
&=
\frac{G^2}{3 r^4} \sum_{i=1}^N \langle m_i^2 \rangle
=
\frac{G^2\mu^2}{3 r^4} N\,,
\end{aligned}
\end{equation}
where $\langle x_i x_j \rangle = r^2\delta_{ij}/3$ was used and
$\mu^2(r) \equiv \int_0^\infty \d m\,m^2\phi(m,r)$.  Over densities within the
shell will persist until the orbits of stars within the shell can
re-randomize the stars.  Therefore, unlike \citet{Chat-Hern-Loeb:02},
we expect these perturbations to be correlated for a time comparable
to $1/\Omega_k(r)$, where $\Omega_k(r)$ is the Keplerian angular
velocity at $r$.  Thus, the auto-correlation function of the
acceleration fluctuations is approximately
\begin{equation}
R^{\Delta r}_{a,T} \equiv \langle a(t) a(t+T) \rangle
\simeq
\langle a^2 \rangle \e^{-\Omega_k^2 T^2/2}
=
\frac{G^2\mu^2}{3 r^4} N \e^{-\Omega_k^2 T^2/2}\,.
\end{equation}
This implies a power spectrum associated with the shell of stars of
\begin{equation}
P^{\Delta r}_{a,\omega}
\simeq
\frac{2(2\pi)^{3/2}}{3} \frac{G^2\mu^2}{r^2} \frac{n}{\Omega_k}
\e^{-\omega^2/2\Omega_k^2}\Delta r\,.
\end{equation}
Since we have already assumed that the positions of individual stars are
uncorrelated (when we assumed Poisson statistics within a shell), the
power spectrum of the entire distribution of stars may then be 
constructed by summing multiple shells, i.e.,
\begin{equation}
P_{a,\omega}
=
\int_0^\infty \d r \, 
\frac{2(2\pi)^{3/2}}{3} \frac{G^2\mu^2}{r^2} \frac{n}{\Omega_k}
\e^{-\omega^2/2\Omega_k^2}\,,
\end{equation}
which may be inserted into Equation (\ref{eq:Pxw}) to obtain the
desired power spectrum.

To make further progress we must choose $\phi(m,r)$ and $n(r)$.  From
these we may then construct the $\Omega_k$:
\begin{equation}
\Omega_k^2(r) = \frac{G M}{r^3} + \frac{G}{r^3} \int_0^\infty \d r \int_0^\infty \d m
\,m\phi(m,r)n(r)\,,
\label{eq:Ok}
\end{equation}
and thus $P_{x,\omega}$.

\bibliography{lbh.bib}

\begin{thebibliography}{183}
\expandafter\ifx\csname natexlab\endcsname\relax\def\natexlab#1{#1}\fi

\bibitem[{{Acciari} {et~al.}(2009){Acciari}, {Aliu}, {Arlen}, {Bautista},
  {Beilicke}, {Benbow}, {Bradbury}, {Buckley}, {Bugaev}, {Butt}, \&
  et~al.}]{Acci_etal:09}
{Acciari}, V.~A., {et~al.} 2009, Science, 325, 444

\bibitem[{{Alexander} \& {Hopman}(2009)}]{Alex-Hopm:09}
{Alexander}, T., \& {Hopman}, C. 2009, \apj, 697, 1861

\bibitem[{{Amaro-Seoane} \& {Preto}(2010)}]{Amar-Pret:10}
{Amaro-Seoane}, P., \& {Preto}, M. 2010, ArXiv e-prints

\bibitem[{{Anderson} {et~al.}(2004){Anderson}, {Ulvestad}, \&
  {Ho}}]{Ande-Ulve-Ho:04}
{Anderson}, J.~M., {Ulvestad}, J.~S., \& {Ho}, L.~C. 2004, \apj, 603, 42

\bibitem[{{Backer} \& {Sramek}(1999)}]{Back-Sram:99}
{Backer}, D.~C., \& {Sramek}, R.~A. 1999, \apj, 524, 805

\bibitem[{{Bartko} {et~al.}(2009){Bartko}, {Perrin}, {Brandner}, {Straubmeier},
  {Richichi}, {Gillessen}, {Paumard}, {Hippler}, {Eckart}, {Sch{\"o}ller},
  {Eisenhauer}, {Haubois}, {Lenzen}, {Rabien}, {Cl{\'e}net}, {Ramos}, {Thiel},
  {Berger}, {Baumeister}, {Kellner}, {Cassaing}, {B{\"o}hm}, {Hofmann},
  {Gendron}, {Klein}, {Dodds-Eden}, {Houairi}, {Hormuth}, {Gr{\"a}ter},
  {Kervella}, {Naranjo}, {Genzel}, {F{\'e}dou}, {Henning}, {Hamaus}, {Jocou},
  {Neumann}, {Haug}, {Lacour}, {Laun}, {Kolmeder}, {Malbet}, {Rohloff},
  {Pfuhl}, {Perraut}, {Ziegleder}, {Rouan}, {Rousset}, {Amorim}, \&
  {Lima}}]{Bart_etal:09}
{Bartko}, H., {et~al.} 2009, \nar, 53, 301

\bibitem[{{Becker} {et~al.}(1991){Becker}, {White}, \&
  {Edwards}}]{Beck-Whit-Edwa:91}
{Becker}, R.~H., {White}, R.~L., \& {Edwards}, A.~L. 1991, \apjs, 75, 1

\bibitem[{{Begelman} {et~al.}(1980){Begelman}, {Blandford}, \&
  {Rees}}]{Bege-Blan-Rees:80}
{Begelman}, M.~C., {Blandford}, R.~D., \& {Rees}, M.~J. 1980, \nat, 287, 307

\bibitem[{{Bennett} {et~al.}(1986){Bennett}, {Lawrence}, {Burke}, {Hewitt}, \&
  {Mahoney}}]{Benn_etal:86}
{Bennett}, C.~L., {Lawrence}, C.~R., {Burke}, B.~F., {Hewitt}, J.~N., \&
  {Mahoney}, J. 1986, \apjs, 61, 1

\bibitem[{{Biretta} {et~al.}(1999){Biretta}, {Sparks}, \&
  {Macchetto}}]{Bire-Spar-Macc:99}
{Biretta}, J.~A., {Sparks}, W.~B., \& {Macchetto}, F. 1999, \apj, 520, 621

\bibitem[{{Boehm} {et~al.}(2006{\natexlab{a}}){Boehm}, {Niell}, {Tregoning}, \&
  {Schuh}}]{Boeh-Neil-Treg-Schu:06}
{Boehm}, J., {Niell}, A., {Tregoning}, P., \& {Schuh}, H. 2006{\natexlab{a}},
  \grl, 33, 7304

\bibitem[{{Boehm} {et~al.}(2006{\natexlab{b}}){Boehm}, {Werl}, \&
  {Schuh}}]{Boeh-Werl-Schu:06}
{Boehm}, J., {Werl}, B., \& {Schuh}, H. 2006{\natexlab{b}}, Journal of
  Geophysical Research (Solid Earth), 111, 2406

\bibitem[{{Bogdanovi{\'c}} {et~al.}(2009){Bogdanovi{\'c}}, {Eracleous}, \&
  {Sigurdsson}}]{Bogd-Erac-Sigu:08}
{Bogdanovi{\'c}}, T., {Eracleous}, M., \& {Sigurdsson}, S. 2009, \apj, 697, 288

\bibitem[{{Boroson} \& {Lauer}(2009)}]{Boro-Laue:09}
{Boroson}, T.~A., \& {Lauer}, T.~R. 2009, \nat, 458, 53

\bibitem[{{Bower} {et~al.}(2001){Bower}, {Backer}, \&
  {Sramek}}]{Bowe-Back-Sram:01}
{Bower}, G.~C., {Backer}, D.~C., \& {Sramek}, R.~A. 2001, \apj, 558, 127

\bibitem[{{Bower} {et~al.}(1999){Bower}, {Backer}, {Zhao}, {Goss}, \&
  {Falcke}}]{Bowe_etal:99}
{Bower}, G.~C., {Backer}, D.~C., {Zhao}, J., {Goss}, M., \& {Falcke}, H. 1999,
  \apj, 521, 582

\bibitem[{{Bower} {et~al.}(2002){Bower}, {Falcke}, {Sault}, \&
  {Backer}}]{Bowe-Falc-Saul-Back:02}
{Bower}, G.~C., {Falcke}, H., {Sault}, R.~J., \& {Backer}, D.~C. 2002, \apj,
  571, 843

\bibitem[{{Braun} {et~al.}(2007){Braun}, {Oosterloo}, {Morganti}, {Klein}, \&
  {Beck}}]{Brau_etal:07}
{Braun}, R., {Oosterloo}, T.~A., {Morganti}, R., {Klein}, U., \& {Beck}, R.
  2007, \aap, 461, 455

\bibitem[{{Broderick} {et~al.}(2009{\natexlab{a}}){Broderick}, {Fish},
  {Doeleman}, \& {Loeb}}]{Brod_etal:09}
{Broderick}, A.~E., {Fish}, V.~L., {Doeleman}, S.~S., \& {Loeb}, A.
  2009{\natexlab{a}}, \apj, 697, 45

\bibitem[{{Broderick} {et~al.}(2010){Broderick}, {Fish}, {Doeleman}, \&
  {Loeb}}]{Brod_etal:10}
---. 2010, \apj

\bibitem[{{Broderick} \& {Loeb}(2005)}]{Brod-Loeb:05}
{Broderick}, A.~E., \& {Loeb}, A. 2005, \mnras, 363, 353

\bibitem[{{Broderick} \& {Loeb}(2006)}]{Brod-Loeb:06b}
---. 2006, \mnras, 367, 905

\bibitem[{{Broderick} \& {Loeb}(2009)}]{Brod-Loeb:09}
---. 2009, \apj, 697, 1164

\bibitem[{{Broderick} {et~al.}(2009{\natexlab{b}}){Broderick}, {Loeb}, \&
  {Narayan}}]{Brod-Loeb-Nara:09}
{Broderick}, A.~E., {Loeb}, A., \& {Narayan}, R. 2009{\natexlab{b}}, \apj, 701,
  1357

\bibitem[{{Buchholz} {et~al.}(2009){Buchholz}, {Sch{\"o}del}, \&
  {Eckart}}]{Buch_etal:09}
{Buchholz}, R.~M., {Sch{\"o}del}, R., \& {Eckart}, A. 2009, \aap, 499, 483

\bibitem[{{Byun} \& {Bar-Sever}(2009)}]{Byun-BarS:09}
{Byun}, S.~H., \& {Bar-Sever}, Y.~E. 2009, Journal of Geodesy, 83, 367

\bibitem[{{Cara} \& {Lister}(2008)}]{Cara-List:08}
{Cara}, M., \& {Lister}, M.~L. 2008, \apj, 674, 111

\bibitem[{{Caswell} {et~al.}(2010){Caswell}, {Breen}, \&
  {Ellingsen}}]{Casw-Bree-Elli:10}
{Caswell}, J.~L., {Breen}, S.~L., \& {Ellingsen}, S.~P. 2010, \mnras, 1488

\bibitem[{{Chatterjee} {et~al.}(2002){Chatterjee}, {Hernquist}, \&
  {Loeb}}]{Chat-Hern-Loeb:02}
{Chatterjee}, P., {Hernquist}, L., \& {Loeb}, A. 2002, \apj, 572, 371

\bibitem[{{Cohen} {et~al.}(2007){Cohen}, {Lane}, {Cotton}, {Kassim}, {Lazio},
  {Perley}, {Condon}, \& {Erickson}}]{Cohe_etal:07}
{Cohen}, A.~S., {Lane}, W.~M., {Cotton}, W.~D., {Kassim}, N.~E., {Lazio},
  T.~J.~W., {Perley}, R.~A., {Condon}, J.~J., \& {Erickson}, W.~C. 2007, \aj,
  134, 1245

\bibitem[{{Condon} {et~al.}(2002){Condon}, {Cotton}, \&
  {Broderick}}]{Cond-Cott-Brod:02}
{Condon}, J.~J., {Cotton}, W.~D., \& {Broderick}, J.~J. 2002, \aj, 124, 675

\bibitem[{{Condon} {et~al.}(1998){Condon}, {Cotton}, {Greisen}, {Yin},
  {Perley}, {Taylor}, \& {Broderick}}]{Cond_etal:98}
{Condon}, J.~J., {Cotton}, W.~D., {Greisen}, E.~W., {Yin}, Q.~F., {Perley},
  R.~A., {Taylor}, G.~B., \& {Broderick}, J.~J. 1998, \aj, 115, 1693

\bibitem[{{Condon} \& {Jauncey}(1974)}]{Cond-Jaun:74}
{Condon}, J.~J., \& {Jauncey}, D.~L. 1974, \aj, 79, 1220

\bibitem[{{Counselman} {et~al.}(1974){Counselman}, {Kent}, {Knight}, {Shapiro},
  {Clark}, {Hinteregger}, {Rogers}, \& {Whitney}}]{Coun_etal:74}
{Counselman}, III, C.~C., {Kent}, S.~M., {Knight}, C.~A., {Shapiro}, I.~I.,
  {Clark}, T.~A., {Hinteregger}, H.~F., {Rogers}, A.~E.~E., \& {Whitney}, A.~R.
  1974, Physical Review Letters, 33, 1621

\bibitem[{{Covey} {et~al.}(2008){Covey}, {Hawley}, {Bochanski}, {West}, {Reid},
  {Golimowski}, {Davenport}, {Henry}, {Uomoto}, \& {Holtzman}}]{Cove_etal:08}
{Covey}, K.~R., {et~al.} 2008, \aj, 136, 1778

\bibitem[{{Dale} {et~al.}(2005){Dale}, {Bendo}, {Engelbracht}, {Gordon},
  {Regan}, {Armus}, {Cannon}, {Calzetti}, {Draine}, {Helou}, {Joseph},
  {Kennicutt}, {Li}, {Murphy}, {Roussel}, {Walter}, {Hanson}, {Hollenbach},
  {Jarrett}, {Kewley}, {Lamanna}, {Leitherer}, {Meyer}, {Rieke}, {Rieke},
  {Sheth}, {Smith}, \& {Thornley}}]{Dale_etal:05}
{Dale}, D.~A., {et~al.} 2005, \apj, 633, 857

\bibitem[{{Decarli} {et~al.}(2010){Decarli}, {Dotti}, {Montuori}, {Liimets}, \&
  {Ederoclite}}]{Deca_etal:10}
{Decarli}, R., {Dotti}, M., {Montuori}, C., {Liimets}, T., \& {Ederoclite}, A.
  2010, \apjl, 720, L93

\bibitem[{{Dexter} {et~al.}(2010){Dexter}, {Agol}, {Fragile}, \&
  {McKinney}}]{Dext-Agol-Frag-McKi:10}
{Dexter}, J., {Agol}, E., {Fragile}, P.~C., \& {McKinney}, J.~C. 2010, \apj,
  717, 1092

\bibitem[{{Do} {et~al.}(2009){Do}, {Ghez}, {Morris}, {Lu}, {Matthews}, {Yelda},
  \& {Larkin}}]{Do_etal:09}
{Do}, T., {Ghez}, A.~M., {Morris}, M.~R., {Lu}, J.~R., {Matthews}, K., {Yelda},
  S., \& {Larkin}, J. 2009, \apj, 703, 1323

\bibitem[{{Do} {et~al.}(2010){Do}, {Ghez}, {Morris}, {Lu}, {Matthews}, {Yelda},
  {Wright}, \& {Larkin}}]{Do_etal:10}
{Do}, T., {Ghez}, A.~M., {Morris}, M.~R., {Lu}, J.~R., {Matthews}, K., {Yelda},
  S., {Wright}, S., \& {Larkin}, J. 2010, ArXiv e-prints

\bibitem[{{Doeleman} {et~al.}(2009{\natexlab{a}}){Doeleman}, {Agol}, {Backer},
  {Baganoff}, {Bower}, {Broderick}, {Fabian}, {Fish}, {Gammie}, {Ho}, {Honman},
  {Krichbaum}, {Loeb}, {Marrone}, {Reid}, {Rogers}, {Shapiro}, {Strittmatter},
  {Tilanus}, {Weintroub}, {Whitney}, {Wright}, \& {Ziurys}}]{Doel_etal:2010}
{Doeleman}, S., {et~al.} 2009{\natexlab{a}}, in Astronomy, Vol. 2010,
  astro2010: The Astronomy and Astrophysics Decadal Survey, 68--+

\bibitem[{{Doeleman} {et~al.}(2009{\natexlab{b}}){Doeleman}, {Fish},
  {Broderick}, {Loeb}, \& {Rogers}}]{Doel_etal:09}
{Doeleman}, S.~S., {Fish}, V.~L., {Broderick}, A.~E., {Loeb}, A., \& {Rogers},
  A.~E.~E. 2009{\natexlab{b}}, \apj, 695, 59

\bibitem[{{Doeleman} {et~al.}(2008){Doeleman}, {Weintroub}, {Rogers},
  {Plambeck}, {Freund}, {Tilanus}, {Friberg}, {Ziurys}, {Moran}, {Corey},
  {Young}, {Smythe}, {Titus}, {Marrone}, {Cappallo}, {Bock}, {Bower},
  {Chamberlin}, {Davis}, {Krichbaum}, {Lamb}, {Maness}, {Niell}, {Roy},
  {Strittmatter}, {Werthimer}, {Whitney}, \& {Woody}}]{Doel_etal:08}
{Doeleman}, S.~S., {et~al.} 2008, \nat, 455, 78

\bibitem[{{Doi} {et~al.}(2005){Doi}, {Kameno}, {Kohno}, {Nakanishi}, \&
  {Inoue}}]{Doi_etal:05}
{Doi}, A., {Kameno}, S., {Kohno}, K., {Nakanishi}, K., \& {Inoue}, M. 2005,
  \mnras, 363, 692

\bibitem[{{Douglas} {et~al.}(1996){Douglas}, {Bash}, {Bozyan}, {Torrence}, \&
  {Wolfe}}]{Doug_etal:96}
{Douglas}, J.~N., {Bash}, F.~N., {Bozyan}, F.~A., {Torrence}, G.~W., \&
  {Wolfe}, C. 1996, \aj, 111, 1945

\bibitem[{{Dressel} \& {Condon}(1978)}]{Dres-Cond:78}
{Dressel}, L.~L., \& {Condon}, J.~J. 1978, \apjs, 36, 53

\bibitem[{{Eckart} {et~al.}(2006){Eckart}, {Baganoff}, {Sch{\"o}del}, {Morris},
  {Genzel}, {Bower}, {Marrone}, {Moran}, {Viehmann}, {Bautz}, {Brandt},
  {Garmire}, {Ott}, {Trippe}, {Ricker}, {Straubmeier}, {Roberts},
  {Yusef-Zadeh}, {Zhao}, \& {Rao}}]{Ecka_etal:06}
{Eckart}, A., {et~al.} 2006, \aap, 450, 535

\bibitem[{{Ekers}(1969)}]{Eker:69}
{Ekers}, J.~A. 1969, Australian Journal of Physics Astrophysical Supplement, 7,
  3

\bibitem[{{Escala} {et~al.}(2004){Escala}, {Larson}, {Coppi}, \&
  {Mardones}}]{Esca_etal:04}
{Escala}, A., {Larson}, R.~B., {Coppi}, P.~S., \& {Mardones}, D. 2004, \apj,
  607, 765

\bibitem[{{Escala} {et~al.}(2005){Escala}, {Larson}, {Coppi}, \&
  {Mardones}}]{Esca_etal:05}
---. 2005, \apj, 630, 152

\bibitem[{{Filho} {et~al.}(2004){Filho}, {Fraternali}, {Markoff}, {Nagar},
  {Barthel}, {Ho}, \& {Yuan}}]{Filh_etal:04}
{Filho}, M.~E., {Fraternali}, F., {Markoff}, S., {Nagar}, N.~M., {Barthel},
  P.~D., {Ho}, L.~C., \& {Yuan}, F. 2004, \aap, 418, 429

\bibitem[{{Fish} {et~al.}(2009){Fish}, {Doeleman}, {Broderick}, {Loeb}, \&
  {Rogers}}]{Fish_etal:09}
{Fish}, V.~L., {Doeleman}, S.~S., {Broderick}, A.~E., {Loeb}, A., \& {Rogers},
  A.~E.~E. 2009, \apj, 706, 1353

\bibitem[{{Fish} {et~al.}(2011){Fish}, {Doeleman}, {Beaudoin}, {Blundell},
  {Bolin}, {Bower}, {Chamberlin}, {Freund}, {Friberg}, {Gurwell}, {Honma},
  {Inoue}, {Krichbaum}, {Lamb}, {Marrone}, {Moran}, {Oyama}, {Plambeck},
  {Primiani}, {Rogers}, {Smythe}, {SooHoo}, {Strittmatter}, {Tilanus}, {Titus},
  {Weintroub}, {Wright}, {Woody}, {Young}, \& {Ziurys}}]{Fish_etal:10}
{Fish}, V.~L., {et~al.} 2011, \apjl, 727, L36

\bibitem[{{Flesch}(2010)}]{Fles:10}
{Flesch}, E. 2010, \pasa, 27, 283

\bibitem[{{Fomalont}(2005)}]{Foma:05}
{Fomalont}, E.~B. 2005, in Astronomical Society of the Pacific Conference
  Series, Vol. 340, Future Directions in High Resolution Astronomy, ed.
  {J.~Romney \& M.~Reid}, 460--+

\bibitem[{{Freitag} {et~al.}(2006){Freitag}, {Amaro-Seoane}, \&
  {Kalogera}}]{Frei-Amar-Kalo:06}
{Freitag}, M., {Amaro-Seoane}, P., \& {Kalogera}, V. 2006, \apj, 649, 91

\bibitem[{{Fujii} {et~al.}(2009){Fujii}, {Iwasawa}, {Funato}, \&
  {Makino}}]{Fuji-Iwas-Funa-Maki:09}
{Fujii}, M., {Iwasawa}, M., {Funato}, Y., \& {Makino}, J. 2009, \apj, 695, 1421

\bibitem[{{Fujii} {et~al.}(2010){Fujii}, {Iwasawa}, {Funato}, \&
  {Makino}}]{Fuji-Iwas-Funa-Maki:10}
---. 2010, \apjl, 716, L80

\bibitem[{{Gair} {et~al.}(2004){Gair}, {Barack}, {Creighton}, {Cutler},
  {Larson}, {Phinney}, \& {Vallisneri}}]{Gair_etal:04}
{Gair}, J.~R., {Barack}, L., {Creighton}, T., {Cutler}, C., {Larson}, S.~L.,
  {Phinney}, E.~S., \& {Vallisneri}, M. 2004, Classical and Quantum Gravity,
  21, 1595

\bibitem[{{Gair} {et~al.}(2010){Gair}, {Tang}, \&
  {Volonteri}}]{Gair-Tang-Volo:10}
{Gair}, J.~R., {Tang}, C., \& {Volonteri}, M. 2010, \prd, 81, 104014

\bibitem[{{Gallimore} {et~al.}(2006){Gallimore}, {Axon}, {O'Dea}, {Baum}, \&
  {Pedlar}}]{Gall_etal:06}
{Gallimore}, J.~F., {Axon}, D.~J., {O'Dea}, C.~P., {Baum}, S.~A., \& {Pedlar},
  A. 2006, \aj, 132, 546

\bibitem[{{Gebhardt} {et~al.}(2011){Gebhardt}, {Adams}, {Richstone}, {Lauer},
  {Faber}, {Gultekin}, {Murphy}, \& {Tremaine}}]{Gebh_etal:11}
{Gebhardt}, K., {Adams}, J., {Richstone}, D., {Lauer}, T.~R., {Faber}, S.~M.,
  {Gultekin}, K., {Murphy}, J., \& {Tremaine}, S. 2011, ArXiv:1101.1954

\bibitem[{{Gebhardt} \& {Thomas}(2009)}]{Gebh-Thom:09}
{Gebhardt}, K., \& {Thomas}, J. 2009, \apj, 700, 1690

\bibitem[{{Geldzahler} \& {Witzel}(1981)}]{Geld-Witz:81}
{Geldzahler}, B.~J., \& {Witzel}, A. 1981, \aj, 86, 1306

\bibitem[{{Gendre} \& {Wall}(2008)}]{Gend-Wall:08}
{Gendre}, M.~A., \& {Wall}, J.~V. 2008, \mnras, 390, 819

\bibitem[{{Genzel} {et~al.}(1976){Genzel}, {Pauliny-Toth}, {Preuss}, \&
  {Witzel}}]{Genz_etal:76}
{Genzel}, R., {Pauliny-Toth}, I.~I.~K., {Preuss}, E., \& {Witzel}, A. 1976,
  \aj, 81, 1084

\bibitem[{{Genzel} {et~al.}(2003){Genzel}, {Sch{\"o}del}, {Ott}, {Eisenhauer},
  {Hofmann}, {Lehnert}, {Eckart}, {Alexander}, {Sternberg}, {Lenzen},
  {Cl{\'e}net}, {Lacombe}, {Rouan}, {Renzini}, \&
  {Tacconi-Garman}}]{Genz_etal:03}
{Genzel}, R., {et~al.} 2003, \apj, 594, 812

\bibitem[{{Gerhard}(2001)}]{Gerh:01}
{Gerhard}, O. 2001, \apjl, 546, L39

\bibitem[{{Ghez} {et~al.}(2003){Ghez}, {Duch{\^e}ne}, {Matthews}, {Hornstein},
  {Tanner}, {Larkin}, {Morris}, {Becklin}, {Salim}, {Kremenek}, {Thompson},
  {Soifer}, {Neugebauer}, \& {McLean}}]{Ghez_etal:03}
{Ghez}, A.~M., {et~al.} 2003, \apjl, 586, L127

\bibitem[{{Ghez} {et~al.}(2008){Ghez}, {Salim}, {Weinberg}, {Lu}, {Do}, {Dunn},
  {Matthews}, {Morris}, {Yelda}, {Becklin}, {Kremenek}, {Milosavljevic}, \&
  {Naiman}}]{Ghez_etal:08}
---. 2008, \apj, 689, 1044

\bibitem[{{Giannios} {et~al.}(2010){Giannios}, {Uzdensky}, \&
  {Begelman}}]{Gian-Uzde-Bege:10}
{Giannios}, D., {Uzdensky}, D.~A., \& {Begelman}, M.~C. 2010, \mnras, 402, 1649

\bibitem[{{Gillessen} {et~al.}(2009{\natexlab{a}}){Gillessen}, {Eisenhauer},
  {Fritz}, {Bartko}, {Dodds-Eden}, {Pfuhl}, {Ott}, \& {Genzel}}]{Gill_etal:09b}
{Gillessen}, S., {Eisenhauer}, F., {Fritz}, T.~K., {Bartko}, H., {Dodds-Eden},
  K., {Pfuhl}, O., {Ott}, T., \& {Genzel}, R. 2009{\natexlab{a}}, \apjl, 707,
  L114

\bibitem[{{Gillessen} {et~al.}(2009{\natexlab{b}}){Gillessen}, {Eisenhauer},
  {Trippe}, {Alexander}, {Genzel}, {Martins}, \& {Ott}}]{Gill_etal:09a}
{Gillessen}, S., {Eisenhauer}, F., {Trippe}, S., {Alexander}, T., {Genzel}, R.,
  {Martins}, F., \& {Ott}, T. 2009{\natexlab{b}}, \apj, 692, 1075

\bibitem[{{Gower} {et~al.}(1967){Gower}, {Scott}, \&
  {Wills}}]{Gowe-Scot-Will:67}
{Gower}, J.~F.~R., {Scott}, P.~F., \& {Wills}, D. 1967, \memras, 71, 49

\bibitem[{{Gregory} \& {Condon}(1991)}]{Greg-Cond:91}
{Gregory}, P.~C., \& {Condon}, J.~J. 1991, \apjs, 75, 1011

\bibitem[{{Gregory} {et~al.}(1996){Gregory}, {Scott}, {Douglas}, \&
  {Condon}}]{Greg-Scot-Doug-Cond:96}
{Gregory}, P.~C., {Scott}, W.~K., {Douglas}, K., \& {Condon}, J.~J. 1996,
  \apjs, 103, 427

\bibitem[{{Gualandris} \& {Merritt}(2009)}]{Gual-Merr:09}
{Gualandris}, A., \& {Merritt}, D. 2009, \apj, 705, 361

\bibitem[{{Hansen} \& {Milosavljevi{\'c}}(2003)}]{Hans-Milo:03}
{Hansen}, B.~M.~S., \& {Milosavljevi{\'c}}, M. 2003, \apjl, 593, L77

\bibitem[{{Healey} {et~al.}(2007){Healey}, {Romani}, {Taylor}, {Sadler},
  {Ricci}, {Murphy}, {Ulvestad}, \& {Winn}}]{Heal_etal:07}
{Healey}, S.~E., {Romani}, R.~W., {Taylor}, G.~B., {Sadler}, E.~M., {Ricci},
  R., {Murphy}, T., {Ulvestad}, J.~S., \& {Winn}, J.~N. 2007, \apjs, 171, 61

\bibitem[{{Heeschen} \& {Conklin}(1975)}]{Hees-Conk:75}
{Heeschen}, D.~S., \& {Conklin}, E.~K. 1975, \apj, 196, 347

\bibitem[{{Heeschen} \& {Wade}(1964)}]{Hees-Wade:64}
{Heeschen}, D.~S., \& {Wade}, C.~M. 1964, \aj, 69, 277

\bibitem[{{Honma} {et~al.}(2000){Honma}, {Kawaguchi}, \&
  {Sasao}}]{Honm-Kawa-Sasa:00}
{Honma}, M., {Kawaguchi}, N., \& {Sasao}, T. 2000, in Society of Photo-Optical
  Instrumentation Engineers (SPIE) Conference Series, Vol. 4015, Society of
  Photo-Optical Instrumentation Engineers (SPIE) Conference Series, ed.
  {H.~R.~Butcher}, 624--631

\bibitem[{{Hopman} \& {Alexander}(2005)}]{Hopm-Alex:05}
{Hopman}, C., \& {Alexander}, T. 2005, \apj, 629, 362

\bibitem[{{Hopman} \& {Alexander}(2006)}]{Hopm-Alex:06}
---. 2006, \apjl, 645, L133

\bibitem[{{Horiuchi} {et~al.}(2004){Horiuchi}, {Fomalont}, {Taylor}, {Scott},
  {Lovell}, {Moellenbrock}, {Dodson}, {Murata}, {Hirabayashi}, {Edwards},
  {Gurvits}, \& {Shen}}]{Hori_etal:04}
{Horiuchi}, S., {et~al.} 2004, \apj, 616, 110

\bibitem[{{Huang} {et~al.}(2009){Huang}, {Takahashi}, \&
  {Shen}}]{Huan-Taka-Shen:09}
{Huang}, L., {Takahashi}, R., \& {Shen}, Z. 2009, \apj, 706, 960

\bibitem[{{Humphreys}(2007)}]{Hump:07}
{Humphreys}, E.~M.~L. 2007, in IAU Symposium, Vol. 242, IAU Symposium, ed.
  {J.~M.~Chapman \& W.~A.~Baan}, 471--480

\bibitem[{{Imai} {et~al.}(1997){Imai}, {Sasao}, {Kameya}, {Miyoshi}, {Shibata},
  {Asaki}, {Omodaka}, {Morimoto}, {Mochizuki}, {Suzuyama}, {Iguchi}, {Kameno},
  {Jike}, {Iwadate}, {Sakai}, {Miyaji}, {Kawaguchi}, \&
  {Miyazawa}}]{Imai_etal:97}
{Imai}, H., {et~al.} 1997, \aap, 317, L67

\bibitem[{{Isaacman} {et~al.}(1980){Isaacman}, {Wouterloot}, \&
  {Habing}}]{Isaa-Wout-Habi:80}
{Isaacman}, R., {Wouterloot}, J.~G.~A., \& {Habing}, H.~J. 1980, \aap, 86, 254

\bibitem[{{Junor} {et~al.}(1999){Junor}, {Biretta}, \&
  {Livio}}]{Juno-Bire-Livi:99}
{Junor}, W., {Biretta}, J.~A., \& {Livio}, M. 1999, \nat, 401, 891

\bibitem[{{Kawaguchi} {et~al.}(2000){Kawaguchi}, {Sasao}, \&
  {Manabe}}]{Kawa-Sasa-Mana:00}
{Kawaguchi}, N., {Sasao}, T., \& {Manabe}, S. 2000, in Society of Photo-Optical
  Instrumentation Engineers (SPIE) Conference Series, Vol. 4015, Society of
  Photo-Optical Instrumentation Engineers (SPIE) Conference Series, ed.
  {H.~R.~Butcher}, 544--551

\bibitem[{{Kellermann} \& {Pauliny-Toth}(1973)}]{Kell-Paul:73}
{Kellermann}, K.~I., \& {Pauliny-Toth}, I.~I.~K. 1973, \aj, 78, 828

\bibitem[{{Kellermann} {et~al.}(1969){Kellermann}, {Pauliny-Toth}, \&
  {Williams}}]{Kell-Paul-Will:69}
{Kellermann}, K.~I., {Pauliny-Toth}, I.~I.~K., \& {Williams}, P.~J.~S. 1969,
  \apj, 157, 1

\bibitem[{{Kellermann} {et~al.}(2004){Kellermann}, {Lister}, {Homan},
  {Vermeulen}, {Cohen}, {Ros}, {Kadler}, {Zensus}, \& {Kovalev}}]{Kell_etal:04}
{Kellermann}, K.~I., {et~al.} 2004, \apj, 609, 539

\bibitem[{{Kim} {et~al.}(2004){Kim}, {Figer}, \& {Morris}}]{Kim-Fige-Morr:04}
{Kim}, S.~S., {Figer}, D.~F., \& {Morris}, M. 2004, \apjl, 607, L123

\bibitem[{{Kovalev} {et~al.}(2007){Kovalev}, {Lister}, {Homan}, \&
  {Kellermann}}]{Kova-List-Homa-Kell:07}
{Kovalev}, Y.~Y., {Lister}, M.~L., {Homan}, D.~C., \& {Kellermann}, K.~I. 2007,
  \apjl, 668, L27

\bibitem[{{Kovalev} {et~al.}(2005){Kovalev}, {Kellermann}, {Lister}, {Homan},
  {Vermeulen}, {Cohen}, {Ros}, {Kadler}, {Lobanov}, {Zensus}, {Kardashev},
  {Gurvits}, {Aller}, \& {Aller}}]{Kova_etal:05}
{Kovalev}, Y.~Y., {et~al.} 2005, \aj, 130, 2473

\bibitem[{{Krichbaum} {et~al.}(2006){Krichbaum}, {Graham}, {Bremer}, {Alef},
  {Witzel}, {Zensus}, \& {Eckart}}]{Kric_etal:06}
{Krichbaum}, T.~P., {Graham}, D.~A., {Bremer}, M., {Alef}, W., {Witzel}, A.,
  {Zensus}, J.~A., \& {Eckart}, A. 2006, J. Phys. Conf. Series, 54, 328

\bibitem[{{Krichbaum} {et~al.}(1997){Krichbaum}, {Graham}, {Greve}, {Wink},
  {Alcolea}, {Colomer}, {de Vicente}, {Baudry}, {Gomez-Gonzalez}, {Grewing}, \&
  {Witzel}}]{Kric_etal:97}
{Krichbaum}, T.~P., {et~al.} 1997, \aap, 323, L17

\bibitem[{{Krips} {et~al.}(2007){Krips}, {Eckart}, {Krichbaum}, {Pott}, {Leon},
  {Neri}, {Garc{\'{\i}}a-Burillo}, {Combes}, {Boone}, {Baker}, {Tacconi},
  {Schinnerer}, \& {Hunt}}]{Krip_etal:07}
{Krips}, M., {et~al.} 2007, \aap, 464, 553

\bibitem[{{Kuehr} {et~al.}(1981){Kuehr}, {Witzel}, {Pauliny-Toth}, \&
  {Nauber}}]{Kueh_etal:81}
{Kuehr}, H., {Witzel}, A., {Pauliny-Toth}, I.~I.~K., \& {Nauber}, U. 1981,
  \aaps, 45, 367

\bibitem[{{Large} {et~al.}(1981){Large}, {Mills}, {Little}, {Crawford}, \&
  {Sutton}}]{Larg_etal:81}
{Large}, M.~I., {Mills}, B.~Y., {Little}, A.~G., {Crawford}, D.~F., \&
  {Sutton}, J.~M. 1981, \mnras, 194, 693

\bibitem[{{Laurent-Muehleisen} {et~al.}(1997){Laurent-Muehleisen}, {Kollgaard},
  {Ryan}, {Feigelson}, {Brinkmann}, \& {Siebert}}]{Laur_etal:97}
{Laurent-Muehleisen}, S.~A., {Kollgaard}, R.~I., {Ryan}, P.~J., {Feigelson},
  E.~D., {Brinkmann}, W., \& {Siebert}, J. 1997, \aaps, 122, 235

\bibitem[{{Ledlow} \& {Owen}(1995)}]{Ledl-Owen:95}
{Ledlow}, M.~J., \& {Owen}, F.~N. 1995, \aj, 109, 853

\bibitem[{{Lee} {et~al.}(2008){Lee}, {Lobanov}, {Krichbaum}, {Witzel},
  {Zensus}, {Bremer}, {Greve}, \& {Grewing}}]{Lee_etal:08}
{Lee}, S., {Lobanov}, A.~P., {Krichbaum}, T.~P., {Witzel}, A., {Zensus}, A.,
  {Bremer}, M., {Greve}, A., \& {Grewing}, M. 2008, \aj, 136, 159

\bibitem[{{Leeuw} {et~al.}(2000){Leeuw}, {Sansom}, \&
  {Robson}}]{Leeu-Sans-Robs:00}
{Leeuw}, L.~L., {Sansom}, A.~E., \& {Robson}, E.~I. 2000, \mnras, 311, 683

\bibitem[{{Leeuw} {et~al.}(2004){Leeuw}, {Sansom}, {Robson}, {Haas}, \&
  {Kuno}}]{Leeu_etal:04}
{Leeuw}, L.~L., {Sansom}, A.~E., {Robson}, E.~I., {Haas}, M., \& {Kuno}, N.
  2004, \apj, 612, 837

\bibitem[{{Levin} {et~al.}(2005){Levin}, {Wu}, \& {Thommes}}]{Levi-Wu-Thom:05}
{Levin}, Y., {Wu}, A., \& {Thommes}, E. 2005, \apj, 635, 341

\bibitem[{{Lister} \& {Homan}(2005)}]{List-Homa:05}
{Lister}, M.~L., \& {Homan}, D.~C. 2005, \aj, 130, 1389

\bibitem[{{Liu} {et~al.}(2010){Liu}, {Shen}, {Strauss}, \&
  {Greene}}]{Liu-Shen-Stra-Gree:10}
{Liu}, X., {Shen}, Y., {Strauss}, M.~A., \& {Greene}, J.~E. 2010, \apj, 708,
  427

\bibitem[{{Lonsdale} {et~al.}(1998){Lonsdale}, {Doeleman}, \&
  {Phillips}}]{Lons-Doel-Phil:98}
{Lonsdale}, C.~J., {Doeleman}, S.~S., \& {Phillips}, R.~B. 1998, \aj, 116, 8

\bibitem[{{Ly} {et~al.}(2007){Ly}, {Walker}, \& {Junor}}]{Ly-Walk-Juno:07}
{Ly}, C., {Walker}, R.~C., \& {Junor}, W. 2007, \apj, 660, 200

\bibitem[{{Ly} {et~al.}(2004){Ly}, {Walker}, \& {Wrobel}}]{Ly-Walk-Wrob:04}
{Ly}, C., {Walker}, R.~C., \& {Wrobel}, J.~M. 2004, \aj, 127, 119

\bibitem[{{Marrone} {et~al.}(2008){Marrone}, {Baganoff}, {Morris}, {Moran},
  {Ghez}, {Hornstein}, {Dowell}, {Mu{\~n}oz}, {Bautz}, {Ricker}, {Brandt},
  {Garmire}, {Lu}, {Matthews}, {Zhao}, {Rao}, \& {Bower}}]{Marr_etal:08}
{Marrone}, D.~P., {et~al.} 2008, \apj, 682, 373

\bibitem[{{Marscher} {et~al.}(2008){Marscher}, {Jorstad}, {D'Arcangelo},
  {Smith}, {Williams}, {Larionov}, {Oh}, {Olmstead}, {Aller}, {Aller},
  {McHardy}, {L{\"a}hteenm{\"a}ki}, {Tornikoski}, {Valtaoja}, {Hagen-Thorn},
  {Kopatskaya}, {Gear}, {Tosti}, {Kurtanidze}, {Nikolashvili}, {Sigua},
  {Miller}, \& {Ryle}}]{Mars_etal:08}
{Marscher}, A.~P., {et~al.} 2008, \nat, 452, 966

\bibitem[{{Masson}(1994)}]{Mass:94}
{Masson}, C.~R. 1994, in Astronomical Society of the Pacific Conference Series,
  Vol.~59, IAU Colloq. 140: Astronomy with Millimeter and Submillimeter Wave
  Interferometry, ed. {M.~Ishiguro \& J.~Welch}, 87--+

\bibitem[{{McMillan} \& {Portegies Zwart}(2003)}]{McMi-Port:03}
{McMillan}, S.~L.~W., \& {Portegies Zwart}, S.~F. 2003, \apj, 596, 314

\bibitem[{{Menten} {et~al.}(1997){Menten}, {Reid}, {Eckart}, \&
  {Genzel}}]{Ment_etal:97}
{Menten}, K.~M., {Reid}, M.~J., {Eckart}, A., \& {Genzel}, R. 1997, \apjl, 475,
  L111+

\bibitem[{{Merritt}(2010)}]{Merr:10}
{Merritt}, D. 2010, \apj, 718, 739

\bibitem[{{Merritt} {et~al.}(2009){Merritt}, {Gualandris}, \&
  {Mikkola}}]{Merr-Gual-Mikk:09}
{Merritt}, D., {Gualandris}, A., \& {Mikkola}, S. 2009, \apjl, 693, L35

\bibitem[{{Merritt} \& {Poon}(2004)}]{Merr-Poon:04}
{Merritt}, D., \& {Poon}, M.~Y. 2004, \apj, 606, 788

\bibitem[{{Meyer} {et~al.}(2006){Meyer}, {Sch{\"o}del}, {Eckart}, {Karas},
  {Dov{\v c}iak}, \& {Duschl}}]{Meye_etal:06}
{Meyer}, L., {Sch{\"o}del}, R., {Eckart}, A., {Karas}, V., {Dov{\v c}iak}, M.,
  \& {Duschl}, W.~J. 2006, \aap, 458, L25

\bibitem[{{Miller} {et~al.}(2009){Miller}, {Alexander}, {Amaro-Seoane},
  {Barth}, {Cutler}, {Gaier}, {Hopman}, {Merritt}, {Phinney}, \&
  {Richstone}}]{Mill_etal:09}
{Miller}, M.~C., {et~al.} 2009, in Astronomy, Vol. 2010, astro2010: The
  Astronomy and Astrophysics Decadal Survey, 210--+

\bibitem[{{Milosavljevi{\'c}} \& {Merritt}(2003)}]{Milo-Merr:03}
{Milosavljevi{\'c}}, M., \& {Merritt}, D. 2003, \apj, 596, 860

\bibitem[{{Miralda-Escud{\'e}} \& {Gould}(2000)}]{Mira-Goul:00}
{Miralda-Escud{\'e}}, J., \& {Gould}, A. 2000, \apj, 545, 847

\bibitem[{{Morris}(1993)}]{Morr:93}
{Morris}, M. 1993, \apj, 408, 496

\bibitem[{{Mo{\'s}cibrodzka} {et~al.}(2009){Mo{\'s}cibrodzka}, {Gammie},
  {Dolence}, {Shiokawa}, \& {Leung}}]{Mosc_etal:09}
{Mo{\'s}cibrodzka}, M., {Gammie}, C.~F., {Dolence}, J.~C., {Shiokawa}, H., \&
  {Leung}, P.~K. 2009, \apj, 706, 497

\bibitem[{{Nagar} {et~al.}(2005){Nagar}, {Falcke}, \&
  {Wilson}}]{Naga-Falc-Wils:05}
{Nagar}, N.~M., {Falcke}, H., \& {Wilson}, A.~S. 2005, \aap, 435, 521

\bibitem[{{Nagar} {et~al.}(2002){Nagar}, {Falcke}, {Wilson}, \&
  {Ulvestad}}]{Naga-Falc-Wils-Ulve:02}
{Nagar}, N.~M., {Falcke}, H., {Wilson}, A.~S., \& {Ulvestad}, J.~S. 2002, \aap,
  392, 53

\bibitem[{{Nagar} {et~al.}(2001){Nagar}, {Wilson}, \&
  {Falcke}}]{Naga-Wils-Falc:01}
{Nagar}, N.~M., {Wilson}, A.~S., \& {Falcke}, H. 2001, \apjl, 559, L87

\bibitem[{{Niell}(1996)}]{Niel:96}
{Niell}, A.~E. 1996, \jgr, 101, 3227

\bibitem[{{Nord} {et~al.}(2004){Nord}, {Lazio}, {Kassim}, {Hyman}, {LaRosa},
  {Brogan}, \& {Duric}}]{Nord_etal:04}
{Nord}, M.~E., {Lazio}, T.~J.~W., {Kassim}, N.~E., {Hyman}, S.~D., {LaRosa},
  T.~N., {Brogan}, C.~L., \& {Duric}, N. 2004, \aj, 128, 1646

\bibitem[{NRAO(1998)}]{NRAO:98}
NRAO. 1998, {Recommended Site for the Millimeter Array}, Tech. rep., National
  Radio Astronomy Observatory

\bibitem[{{Ochsenbein} {et~al.}(2000){Ochsenbein}, {Bauer}, \&
  {Marcout}}]{VizieR:00}
{Ochsenbein}, F., {Bauer}, P., \& {Marcout}, J. 2000, \aaps, 143, 23

\bibitem[{{O'Leary} {et~al.}(2009){O'Leary}, {Kocsis}, \&
  {Loeb}}]{OLea-Kocs-Loeb:09}
{O'Leary}, R.~M., {Kocsis}, B., \& {Loeb}, A. 2009, \mnras, 395, 2127

\bibitem[{{Olmi} \& {Downes}(1992)}]{Olmi-Down:92}
{Olmi}, L., \& {Downes}, D. 1992, \aap, 262, 634

\bibitem[{{Pauliny-Toth} \& {Kellermann}(1972)}]{Paul-Kell:72}
{Pauliny-Toth}, I.~I.~K., \& {Kellermann}, K.~I. 1972, \aj, 77, 797

\bibitem[{{Pauliny-Toth} {et~al.}(1966){Pauliny-Toth}, {Wade}, \&
  {Heeschen}}]{Paul-Wade-Hees:66}
{Pauliny-Toth}, I.~I.~K., {Wade}, C.~M., \& {Heeschen}, D.~S. 1966, \apjs, 13,
  65

\bibitem[{{Pauliny-Toth} {et~al.}(1978){Pauliny-Toth}, {Witzel}, {Preuss},
  {K{\"u}hr}, {Kellermann}, {Fomalont}, \& {Davis}}]{Paul_etal:78}
{Pauliny-Toth}, I.~I.~K., {Witzel}, A., {Preuss}, E., {K{\"u}hr}, H.,
  {Kellermann}, K.~I., {Fomalont}, E.~B., \& {Davis}, M.~M. 1978, \aj, 83, 451

\bibitem[{{Perets} \& {Alexander}(2008)}]{Pere-Alex:08}
{Perets}, H.~B., \& {Alexander}, T. 2008, \apj, 677, 146

\bibitem[{{Pilkington} \& {Scott}(1965)}]{Pilk-Scot:65}
{Pilkington}, J.~D.~H., \& {Scott}, P.~F. 1965, \memras, 69, 183

\bibitem[{{Porquet} {et~al.}(2008){Porquet}, {Grosso}, {Predehl}, {Hasinger},
  {Yusef-Zadeh}, {Aschenbach}, {Trap}, {Melia}, {Warwick}, {Goldwurm},
  {B{\'e}langer}, {Tanaka}, {Genzel}, {Dodds-Eden}, {Sakano}, \&
  {Ferrando}}]{Porq_etal:08}
{Porquet}, D., {et~al.} 2008, \aap, 488, 549

\bibitem[{{Qin} {et~al.}(2008){Qin}, {Zhao}, {Moran}, {Marrone}, {Patel},
  {Wang}, {Liu}, \& {Kuan}}]{Qin_etal:08}
{Qin}, S., {Zhao}, J., {Moran}, J.~M., {Marrone}, D.~P., {Patel}, N.~A.,
  {Wang}, J., {Liu}, S., \& {Kuan}, Y. 2008, \apj, 677, 353

\bibitem[{{Reich} {et~al.}(2000){Reich}, {F{\"u}rst}, {Reich}, {Kothes},
  {Brinkmann}, \& {Siebert}}]{Reic_etal:00}
{Reich}, W., {F{\"u}rst}, E., {Reich}, P., {Kothes}, R., {Brinkmann}, W., \&
  {Siebert}, J. 2000, \aap, 363, 141

\bibitem[{{Reid} {et~al.}(1989){Reid}, {Biretta}, {Junor}, {Muxlow}, \&
  {Spencer}}]{Reid_etal:89}
{Reid}, M.~J., {Biretta}, J.~A., {Junor}, W., {Muxlow}, T.~W.~B., \& {Spencer},
  R.~E. 1989, \apj, 336, 112

\bibitem[{{Reid} {et~al.}(2008){Reid}, {Broderick}, {Loeb}, {Honma}, \&
  {Brunthaler}}]{Reid_etal:08}
{Reid}, M.~J., {Broderick}, A.~E., {Loeb}, A., {Honma}, M., \& {Brunthaler}, A.
  2008, \apj, 682, 1041

\bibitem[{{Reid} \& {Brunthaler}(2004)}]{Reid-Brun:04}
{Reid}, M.~J., \& {Brunthaler}, A. 2004, \apj, 616, 872

\bibitem[{{Reid} \& {Brunthaler}(2005)}]{Reid-Brun:05}
{Reid}, M.~J., \& {Brunthaler}, A. 2005, in Astronomical Society of the Pacific
  Conference Series, Vol. 340, Future Directions in High Resolution Astronomy,
  ed. {J.~Romney \& M.~Reid}, 253--+

\bibitem[{{Reid} \& {Menten}(2007)}]{Reid-Ment:07}
{Reid}, M.~J., \& {Menten}, K.~M. 2007, \apj, 671, 2068

\bibitem[{{Reid} {et~al.}(2003){Reid}, {Menten}, {Genzel}, {Ott},
  {Sch{\"o}del}, \& {Eckart}}]{Reid_etal:03}
{Reid}, M.~J., {Menten}, K.~M., {Genzel}, R., {Ott}, T., {Sch{\"o}del}, R., \&
  {Eckart}, A. 2003, \apj, 587, 208

\bibitem[{{Reid} {et~al.}(2007){Reid}, {Menten}, {Trippe}, {Ott}, \&
  {Genzel}}]{Reid_etal:07}
{Reid}, M.~J., {Menten}, K.~M., {Trippe}, S., {Ott}, T., \& {Genzel}, R. 2007,
  \apj, 659, 378

\bibitem[{{Reid} {et~al.}(2009){Reid}, {Menten}, {Zheng}, {Brunthaler}, \&
  {Xu}}]{Reid_etal:09}
{Reid}, M.~J., {Menten}, K.~M., {Zheng}, X.~W., {Brunthaler}, A., \& {Xu}, Y.
  2009, \apj, 705, 1548

\bibitem[{{Reid} \& {Moran}(1981)}]{Reid-Mora:81}
{Reid}, M.~J., \& {Moran}, J.~M. 1981, \araa, 19, 231

\bibitem[{{Reid} {et~al.}(1999{\natexlab{a}}){Reid}, {Readhead}, {Vermeulen},
  \& {Treuhaft}}]{Reid-Read-Verm-Treu:99}
{Reid}, M.~J., {Readhead}, A.~C.~S., {Vermeulen}, R.~C., \& {Treuhaft}, R.~N.
  1999{\natexlab{a}}, \apj, 524, 816

\bibitem[{{Reid} {et~al.}(1999{\natexlab{b}}){Reid}, {Readhead}, {Vermeulen},
  \& {Treuhaft}}]{Reid_etal:99}
---. 1999{\natexlab{b}}, \apj, 524, 816

\bibitem[{{Roos}(1981)}]{Roos:81}
{Roos}, N. 1981, \aap, 104, 218

\bibitem[{{Roy} {et~al.}(2005){Roy}, {Rao}, \&
  {Subrahmanyan}}]{Roy-Rao-Subr:05}
{Roy}, S., {Rao}, A.~P., \& {Subrahmanyan}, R. 2005, \mnras, 360, 1305

\bibitem[{{Sch{\"o}del} {et~al.}(2007){Sch{\"o}del}, {Eckart}, {Alexander},
  {Merritt}, {Genzel}, {Sternberg}, {Meyer}, {Kul}, {Moultaka}, {Ott}, \&
  {Straubmeier}}]{Scho_etal:07}
{Sch{\"o}del}, R., {et~al.} 2007, \aap, 469, 125

\bibitem[{{Sch{\"o}nrich} {et~al.}(2010){Sch{\"o}nrich}, {Binney}, \&
  {Dehnen}}]{Scho-Binn-Dehn:10}
{Sch{\"o}nrich}, R., {Binney}, J., \& {Dehnen}, W. 2010, \mnras, 403, 1829

\bibitem[{{Schutz} {et~al.}(2009){Schutz}, {Centrella}, {Cutler}, \&
  {Hughes}}]{Shut_etal:09}
{Schutz}, B.~F., {Centrella}, J., {Cutler}, C., \& {Hughes}, S.~A. 2009, in
  Astronomy, Vol. 2010, astro2010: The Astronomy and Astrophysics Decadal
  Survey, 265--+

\bibitem[{{Shimmins} {et~al.}(1975){Shimmins}, {Bolton}, \&
  {Wall}}]{Shim-Bolt-Wall:75}
{Shimmins}, A.~J., {Bolton}, J.~G., \& {Wall}, J.~V. 1975, Australian Journal
  of Physics Astrophysical Supplement, 34, 63

\bibitem[{{Sigurdsson} \& {Rees}(1997)}]{Sigu-Rees:97}
{Sigurdsson}, S., \& {Rees}, M.~J. 1997, \mnras, 284, 318

\bibitem[{{Slee}(1995)}]{Slee:95}
{Slee}, O.~B. 1995, Australian Journal of Physics, 48, 143

\bibitem[{{Slee} \& {Higgins}(1973)}]{Slee-Higg:73}
{Slee}, O.~B., \& {Higgins}, C.~S. 1973, Australian Journal of Physics
  Astrophysical Supplement, 27, 1

\bibitem[{{Sramek}(1975)}]{Sram:75}
{Sramek}, R. 1975, \aj, 80, 771

\bibitem[{{Stull}(1971)}]{Stul:71}
{Stull}, M.~A. 1971, \aj, 76, 1

\bibitem[{{Thompson} {et~al.}(2001){Thompson}, {Moran}, \&
  {Swenson}}]{Thom-Mora-Swen:01}
{Thompson}, A.~R., {Moran}, J.~M., \& {Swenson}, Jr., G.~W. 2001,
  {Interferometry and Synthesis in Radio Astronomy, 2nd Edition}, ed.
  {Thompson, A.~R., Moran, J.~M., \& Swenson, G.~W., Jr.}

\bibitem[{{Ulvestad} \& {Ho}(2001)}]{Ulve-Ho:01}
{Ulvestad}, J.~S., \& {Ho}, L.~C. 2001, \apjl, 562, L133

\bibitem[{{V{\'e}ron}(1977)}]{Vero:77}
{V{\'e}ron}, P. 1977, \aaps, 30, 131

\bibitem[{{Veron-Cetty} \& {Veron}(1983)}]{Vero-Vero:83}
{Veron-Cetty}, M.~P., \& {Veron}, P. 1983, \aaps, 53, 219

\bibitem[{{Vollmer} {et~al.}(2004){Vollmer}, {Thierbach}, \&
  {Wielebinski}}]{Voll-Thie-Wiel:04}
{Vollmer}, B., {Thierbach}, M., \& {Wielebinski}, R. 2004, \aap, 418, 1

\bibitem[{{Wagner} {et~al.}(2009){Wagner}, {Beilicke}, {Davies}, {Hardee},
  {Krawczynski}, {Mazin}, {Walker}, {Raue}, {Wagner}, {Ly}, {Junor}, {MAGIC},
  {VERITAS}, \& {H.~E.~S.~S.~collaborations}}]{Wagn_etal:09}
{Wagner}, R.~M., {et~al.} 2009, ArXiv e-prints

\bibitem[{{Walker} {et~al.}(2008){Walker}, {Ly}, {Junor}, \&
  {Hardee}}]{Walk-Ly-Juno-Hard:08}
{Walker}, R.~C., {Ly}, C., {Junor}, W., \& {Hardee}, P.~J. 2008, J. Phys. Conf.
  Series, 131, 012053

\bibitem[{{Weintroub}(2008)}]{Wein:08}
{Weintroub}, J. 2008, Journal of Physics Conference Series, 131, 012047

\bibitem[{{White} \& {Becker}(1992)}]{Whit-Beck:92}
{White}, R.~L., \& {Becker}, R.~H. 1992, \apjs, 79, 331

\bibitem[{{White} {et~al.}(1997){White}, {Becker}, {Helfand}, \&
  {Gregg}}]{Whit-Beck-Helf-Greg:97}
{White}, R.~L., {Becker}, R.~H., {Helfand}, D.~J., \& {Gregg}, M.~D. 1997,
  \apj, 475, 479

\bibitem[{{Wills}(1975)}]{Will:75}
{Wills}, B.~J. 1975, Australian Journal of Physics Astrophysical Supplement,
  38, 1

\bibitem[{{Witzel} {et~al.}(1971){Witzel}, {Veron}, \&
  {Veron}}]{Witz-Vero-Vero:71}
{Witzel}, A., {Veron}, P., \& {Veron}, M.~P. 1971, \aap, 11, 171

\bibitem[{{Woody} {et~al.}(2004){Woody}, {Beasley}, {Bolatto}, {Carlstrom},
  {Harris}, {Hawkins}, {Lamb}, {Looney}, {Mundy}, {Plambeck}, {Scott}, \&
  {Wright}}]{Wood_etal:04}
{Woody}, D.~P., {et~al.} 2004, in Society of Photo-Optical Instrumentation
  Engineers (SPIE) Conference Series, Vol. 5498, Society of Photo-Optical
  Instrumentation Engineers (SPIE) Conference Series, ed. {C.~M.~Bradford,
  P.~A.~R.~Ade, J.~E.~Aguirre, J.~J.~Bock, M.~Dragovan, L.~Duband, L.~Earle,
  J.~Glenn, H.~Matsuhara, B.~J.~Naylor, H.~T.~Nguyen, M.~Yun, \&
  J.~Zmuidzinas}, 30--41

\bibitem[{{Wright} \& {Otrupcek}(1990)}]{Wrig-Otru:90}
{Wright}, A., \& {Otrupcek}, R. 1990, in PKS Catalog (1990), 0--+

\bibitem[{{Yu} \& {Tremaine}(2003)}]{Yu-Trem:03}
{Yu}, Q., \& {Tremaine}, S. 2003, \apj, 599, 1129

\bibitem[{{Zavala} \& {Taylor}(2004)}]{Zava-Tayl:04}
{Zavala}, R.~T., \& {Taylor}, G.~B. 2004, \apj, 612, 749

\bibitem[{{Zoonematkermani} {et~al.}(1990){Zoonematkermani}, {Helfand},
  {Becker}, {White}, \& {Perley}}]{Zoon_etal:90}
{Zoonematkermani}, S., {Helfand}, D.~J., {Becker}, R.~H., {White}, R.~L., \&
  {Perley}, R.~A. 1990, \apjs, 74, 181

\end{thebibliography}
\bibliographystyle{apj}

\end{document}